\documentclass[12pt] {article} 

\usepackage{amsmath,amssymb}

\usepackage[all]{xy}

\usepackage[ansinew]{inputenc}
\usepackage[T1]{fontenc}
\usepackage{lmodern}

\usepackage{times}
\usepackage[]{graphicx}

\numberwithin{equation}{section}

\begin{document}

\title{A construction of full QED using finite dimensional Hilbert space}

\author{Charles Francis}


\maketitle

\begin{abstract}
\flushleft\textit{Context}. While causal perturbation theory and lattice regularization allow treatment of the ultraviolet divergences in qed, they do not resolve the mathematical issues of constructive field theory, or show the validity of qed except as a perturbation theory.\\
\textit{Aims}. To present a mathematically rigorous construction of quantum and classical electrodynamics from fundamental principles of quantum theory.\\
\textit{Methods}. Hilbert space of dimension $N$ is justified from statements about measurements with finite range and resolution. Using linear combinations of basis kets, a continuum of kets, $|x\rangle$ for $x\in \mathbb{R}^3$, is constructed such that the inner product can be expressed either as a finite sum or as an integral. Vectors are smooth wave functions such that differential operators are defined and a form of covariance is obeyed (the choice of basis has no affect on underlying physics). Quantum field operators, $\phi(x)$ for $x\in \mathbb{R}^4$, are constructed from creation and annihilation operators on Fock space, obey quantum covariance and locality, and are suitable for a description of particle interactions under the Feynman-St\"{u}ckelberg interpretation. \\
\textit{Results}. It is shown that the formulation is consistent and that any dependency on a lattice arises from measurement, not from underlying physics. In consequence, and because the continuum is constructed from linear combinations of basis kets, it is not required to take the limit $ N\rightarrow\infty $. Quantum fields are defined on a continuum, and are operator valued functions, not distributions. The interacting Dirac equation, Maxwell's equations and the Lorentz force law are derived, showing that qed is a complete theory of the electromagnetic interaction, not just a perturbation theory, and that bare mass and charge are the physical values. Up to the accuracy of measurement, predictions of perturbation theory are identical to those of standard qed with all loop divergences removed. The standard perturbation expansion is asymptotic to the finite expansion given here.
\end{abstract}

\newpage.

\section{Introduction}
\label{sec:1}
\subsection{Background} 
\label{sec:1.1}
The first calculations giving finite results at any order in perturbative qed were carried out in the late 1940s, largely by Tomonaga \cite{Tomonaga}
, Schwinger \cite{Schwinger}\cite{Schwinger2}
, and Feynman \cite{Feynman2}\cite{Feynman3}
. Although these calculations have successfully been built into rigidly defined renormalisation schemes, a constructive approach to qed, showing that it is a mathematically consistent application of quantum mechanics, has been lacking. Among the problems such an approach must address are the requirement of a positive definite norm for valid probabilities, the indefinability of the equal point multiplication between field operators, loop divergences, the Landau pole, the Dyson instability, and classical electromagnetism in the appropriate correspondence.

This paper reviews the Fock space formulation of quantum electrodynamics in the context of an axiomatic formulation of quantum theory using finite dimensional Hilbert space, based upon the principle described by, e.g., Rovelli \cite{Rovelli1} that \textit{all} measured quantities are relational, not just velocity as in relativity. The use of finite dimensions is motivated by showing the correspondence between Hilbert space and a set of formal conditional clauses describing hypothetical measurement results. Since the range and resolution of measurement are intrinsically finite, Hilbert space is finite dimensional, but a continuum of kets,  $|x\rangle$ for $x\in \mathbb{R}^3$, is defined using linear combinations of basis kets, and it is shown that underlying physics is independent of basis (similarly, 3D space does not depend on the coordinates used to describe it). There are technical advantages in using finite dimensional Hilbert space in that stronger theorems are available and the order of taking limits can be tracked. In certain instances (loop integrals) the order of taking limits is critical to whether the limit exists. 

In standard approaches to quantum field theory one starts with a classical field and then quantises it. Space is thus a fundamental physical concept on which the theory is built. Covariance requires that space must be a continuum, and hence that if a lattice is used the limit of small lattice spacing must be taken. In the present treatment quantum properties are understood to arise precisely because space does \textit{not} appear as a fundamental physical concept. Measurement results are seen as relationships between the matter (or radiation) under study and reference matter used to defined the measurement. As such, they are intrinsically frame dependent.  The use of a lattice reflects the finite range and resolution of measurements of position, not a property of a prior space or spacetime.

The interpretation here follows Dirac and Von Neumann, but goes further than either. Von Neumann described quantum logic as a language which tells us what can be discovered from measurement but he did not \textit{translate} the propositions of quantum logic into English. Here Hilbert space is abstracted from the formal statement of sentences in ordinary language. The principle of superposition is not assumed as part of the structure of Hilbert space, but is \textit{exhibited} as a property of conditional and consequent clauses in a formal language describing possible measurement results. Although a finite basis is used for Hilbert space, discrete functions (describing the ket in a given basis) are embedded in a continuum representation such that differential operators are defined and the Dirac delta has a representation as a smooth function. The continuum equations remove any dependency on a specific measurement apparatus and resolution because they contain embedded within them the solutions for all discrete coordinate systems possible in principal or in practice. Thus it will be shown that, although the theory is superficially not covariant, a new form of covariance, quantum covariance, is obeyed (section \ref{sec:3.2}).

A continuum of field operators, $\phi(x)$ for $x\in \mathbb{R}^4$, is defined, obeying quantum covariance and locality. In spite of discreteness, the theory is invariant under changes of lattice, including rotations, boosts, and those due to improvements to experimental technique. The inner product has both sum and integral forms. It is found that this formulation of quantum mechanics allows a complete construction of qed in which Maxwell's equations and the Lorentz force law are derived in the classical correspondence. The fundamental physical concepts are particles, and Feynman diagrams have a natural interpretation in terms of interactions between particles in the absence of spacetime background. The predictions of perturbative qed are unaltered.

\subsection{Outline} 
\label{sec:1.2}
To say that we have carried out a construction of qed from fundamental principles in quantum mechanics it is necessary to give a full treatment. Inevitably such a treatment will contain much that is already known. Small changes in axiomatic structure necessitate the repetition of otherwise standard demonstrations. Often the results are standard, but the reasoning which leads to them is not. For the sake of completeness, I have put into an appendix any demonstration which is known, and which can be removed from the main text without impairing the flow of the presentation.

Section \ref{sec:2} addresses the mathematical foundations of quantum mechanics using finite dimensional Hilbert space. Section \ref{sec:3}, \textit{Particles}, establishes the requirement for a first order Schr\"{o}dinger equation, considers covariance issues, introduces spin and reviews the photon and the Dirac particle. Section \ref{sec:4}, \textit{Interactions}, introduces the interaction Hamiltonian and the Hamiltonian density. The locality requirement is seen from the perturbation expansion. Section \ref{sec:5}, \textit{Field theory}, defines the Dirac field operator, describes the photon and defines the photon field. Without assuming a Lagrangian or classical law, section \ref{sec:6}, \textit{Electromagnetism}, derives the interacting Dirac equation and establishes Maxwell's equations and the Lorentz force law from the minimal interaction in which a Dirac particle emits or absorbs a photon, thereby showing that the physical mass and coupling constant are equal to their bare values in the low energy limit. Section \ref{sec:7}, \textit{Finite quantum electrodynamics}, defines the Dirac propagator and describes the correspondences with causal perturbation theory \cite{Scharf} 
and with lattice regularisation \cite{Creutz}\cite{Montvay}
. The calculation of Feynman rules is given in appendix J. Section \ref{sec:8}, \textit{Conclusion} summarises the results.

\section{Foundations}\label{sec:2}
\subsection{Reference matter}\label{sec:2.1}
When a human observer seeks to quantify nature, he chooses some particular matter from which to define a reference frame or chooses certain matter from which he builds his experimental apparatus. He then observes a defined relationship between this specially, but arbitrarily, chosen reference matter and whatever matter (or radiation) is the subject of study. Here measurement is distinguished from a simple count of a number of objects, and is defined to mean a count of units of a measured quantity, where the definition of the unit of measurement invokes comparison between some aspect of the subject of measurement and a property of the reference matter used to define the unit of measurement. The division between reference matter and subject matter is present in all measurement and appears as the distinction between particle and apparatus in quantum mechanics, and in the definition of position relative to a reference frame in special relativity.

Inertial reference matter is assumed, where inertial is taken to mean that the effect on motion of contact interactions with other matter is negligible. Alternatively inertial coordinates may be calculated from the reference matter; e.g. a satellite spinning on its axis may be used to determine an inertial reference frame, although it is not itself inertial. This introduces complications in the description, but not complications of a fundamental nature.

\subsection{Coordinates}\label{sec:2.2}
We are particularly interested in measurement of time and position. This is sufficient for the study of many (it has been said all) other physical quantities. For convenience, Cartesian coordinates will be chosen. This simplifies certain formulae, but makes no fundamental difference to the treatment. Any apparatus has a finite resolution and the values written down are triplets of terminating decimals, which can be scaled to integers in units of some bounding resolution. Measured positions are always discrete values, determined by the range and resolution of a measurement apparatus. In practice it is simpler to use an equally spaced lattice, containing a very large number $N$ positions given by decimals terminating at some value beyond the best available resolution of any existing apparatus. Margins of error and measurements at lower resolution can be represented using finite sets of such integers. In practice there is also a bound on magnitude. Without loss of generality the same bound, $\nu \in \mathbb{N}$, is used for each coordinate. Knowledge of the ket at any time is thus restricted to this set of triplets and the results of measurement of position are in a (subset of a) finite region, $\mathrm{D} \subset (\chi \mathbb{Z})^3$.\\
\\
\textbf{Postulate:} The \textbf{discrete space coordinate system} is $\mathrm{D}\equiv (-\chi\nu,\chi\nu]^3 \subset (\chi \mathbb{Z})^3$ for some $\nu \in \mathbb{N}$, and for some lattice spacing $\chi \in \mathbb{Q}$  with $\chi>0$.\\

Let $\mathrm{T} \subset \chi \mathbb{Z}$ be a finite discrete time interval such that any particle under study will be measured in $\mathrm{D}$ for times $t \in \mathrm{T}$.\\
\\
\textbf{Postulate:} The \textbf{discrete spacetime coordinate system} is $\mathrm{S} \equiv \mathrm{T} \otimes \mathrm{D}$ and is calibrated such that the speed of light is 1 radially to the origin.\\

The coordinate system is a lattice determined by practical considerations. Not every element of $\mathrm{D}$ need correspond to a possible measurement result, but $\mathrm{D}$ contains as elements or as subsets the possible measurement results for a measurement of position with the chosen apparatus. There is no significance in the bound, $\nu$, of a given coordinate system. It is not intended to take either the limit $\nu \rightarrow \infty$ or $\chi \rightarrow 0$, but $\chi\nu$ is large enough to neglect the possibility of particles leaving $\mathrm{S}$. In practice this is always the case since data is discarded from any trial in which there is not both a well defined initial and final ket; the probability amplitudes defined below relate to conditional probabilities such that both initial and final kets are unambiguously determined (hence there is no detection loophole in Bell tests --- in the absence of unambiguous detection this model does not apply).

\subsection{Particles}\label{sec:2.3}
It is sometimes assumed that a particle is localised in space, even if at unknown location. This is not the case here, since a value of position is not assumed to exist between measurements.\\
\\
\textbf{Postulate:} A \textbf{particle} is any physical entity whose position can be measured at given time such that the result of such measurement is a value, $x \in \mathrm{D}$ , or a neighbourhood $\{x\}\subset \mathrm{D}$ of negligible size.\\
\\
\textbf{Postulate:} An \textbf{elementary particle} is one which cannot, even in principle, be subdivided into particles for which separate positions can be measured.\\
\\
It is not necessary to assume the existence of an elementary particle on metaphysical grounds. If there is such a thing as an elementary particle, then its theoretical properties may be determined, and if something in nature exhibits precisely those properties, then we will claim that it is an elementary particle. Quarks may be considered as elementary particles having separate positions in principle, but bound in practice.

\subsection{Many valued logic}\label{sec:2.4}
Classical logic applies to sets of statements about the real world which are definitely true or definitely false. For example, when we make a statement,\\
\indent\indent$\mathcal{P}(x)$: \textit{The position of a particle is} $x$,\\
we tend to assume that it is definitely true or definitely false. Such statements are said to be \textbf{sharp} or \textbf{crisp}, meaning that they have \textbf{Boolean truth values}, taken from the set $\{0,1\}$. If it is the case that $\mathcal{P}(x)$ is definitely either true or false then classical logic and classical mechanics apply. Similarly, probability theory gives \textbf{Bayesian truth values} from the continuous interval $[0, 1]$ to sentences in the future tense:\\
\indent\indent$\mathcal{Q}(x)$: \textit{In a measurement of position, the result will be} $x$.\\
In quantum mechanics we deal with situations in which there has been no measurement and there is not going to be one. $\mathcal{P}(x)$ and $\mathcal{Q}(x)$ are not then propositions about physical reality. For example, we only get interference from Young's slits when there is no way to determine which slit the particle came through. In the absence of measurement we can consider propositions describing hypothetical measurement results, such as the set of propositions of the form:\\
\indent\indent$\mathcal{R}(x)$: \textit{In a measurement of position, the result would be} $x$.\\
$\mathcal{R}(x)$ is intuitively sensible, even when no measurement is done, but cannot sensibly be given a crisp truth value. Its truth is distinguished from that of $\mathcal{Q}(x)$ because, when no measurements are to be done, we cannot sensibly discuss the potential frequency of individual measurement results.

Quantum logic  \cite{Rescher} 
was introduced by Garrett Birkhoff and John Von Neumann \cite{Birkhoff} 
and is sometimes described as applying counter-intuitive truth values to simple propositions. Here I consider a natural formalisation of statements about hypothetical measurements. Kets are interpreted as formal conditional clauses, rather than as propositions. The dual space consists of corresponding consequent clauses. The inner product combines clauses to generate formal propositions in the subjunctive mood, showing that the language is a consistent and intuitive extension of two-valued logic and classical probability theory. The principle of superposition is simply logical disjunction in formal language; there is no suggestion of an ontological quantity of magnitude $\lvert\langle x|f\rangle\rvert $ associated with a particular particle.

\subsection{Formal language}\label{sec:2.5}
In quantum theory we are not always going to do a measurement, but we want to talk about what \textit{would} happen if we \textit{were} to do a measurement, i.e. we need to be able to make statements about measurement results in the subjunctive mood. I will do this by formally defining kets as conditional clauses. \textbf{Basic conditional clauses}, on which the language is built, refer to individual measurements of position:\\
\\
\textbf{RULE I.} For $x\in\mathrm{D}$, $|x\rangle$  is the \textbf{formal conditional clause} \textit{``If measured position at time $t$ were $x$, \ldots''}.\\

An actual position found by a real apparatus is described by a set of points in the lattice. To describe this we need to extend the language, by introducing an operator corresponding to \textsc{or}, represented by the symbol $+$. To express the idea that one possibility is more likely than the other, we introduce a weighting; thus, if the magnitude of $a$ is greater than that of $b$, then $a|f\rangle + b|g\rangle$ will mean \textit{``if measured position were either $x$ or $y$, but more likely $x$, \ldots''} We also want to be able to express many possibilities, \textit{``If the particle were found at $ x $ or $ y $ or $ z $ or \ldots''}. This is done recursively in rule II:\\
\\
\textbf{RULE II.} If $|g\rangle$ and $|f\rangle$ are formal conditional clauses, and $a$ and $b$ are complex numbers, then $a|g\rangle + b|f\rangle$ is a formal conditional clause. \\

The set of formal conditional clauses, or kets, now has the mathematical structure of an $N$-dimensional vector space, $\mathbb{H}^1(t)$, where $N=8\nu^3$. The elements of $\mathbb{H}^1(t)$ are formal conditional clauses concerning the measurement of position of a single particle at time $t$. Basic conditional clauses, $|x\rangle$, are a basis for $\mathbb{H}^1(t)$. Kets are not strictly states of a particle, but formal conditional clauses describing hypothetical measurement results. They will be referred to as ``states'', in keeping with common practice when no confusion arises. The use of a vector space over the complex numbers (rather than the reals) introduces a degree of freedom which will be used in the description of the evolution of kets.

To complete a formal sentence we need to put a formal conditional clause together with a formal consequent clause. Formal consequent causes refer to a second measurement, at the same time as the first measurement. To make statements about real measurement results we will also need to know how kets evolve in time, but in the first instance the discussion is restricted to hypothetical measurements at time $t$. There is no fundamental difference between one measurement and another, so the grammatical structure, \textit{weighted disjunction}, described in rule II, applies equally well to consequent clauses. These also form an $N$-dimensional vector space, defined from a basis of consequent clauses in one-one correspondence with the basic conditional clauses, or kets, described by rule I. Consequent clauses are represented symbolically by bras:\\
\\
\textbf{RULE III.} $\langle x|$ is the \textbf{formal consequent clause} \textit{``\ldots, then, in a second measurement at time $ t $, measured position would be $x$''}.\\
\\
We put the two clauses together, to make a braket, representing a statement about measurement at a given time:\\
\\
\textbf{RULE IV.} $\langle x|y\rangle$ is the statement \textit{``If measured position at time $ t $ were $ y $, then, in a second measurement at time $ t $, measured position would be $ x $''}.\\

From observation we know that, if, at some particular time, a particle is measured at position $x$, then its position is definitely $x$ and it cannot be measured separately at some other position $y$ at the same time. The statement $\langle x|y\rangle$ is strictly true or false, depending on whether or not $x = y$:\\
\\ 
\textbf{Postulate:} The \textbf{truth value} of $\langle x|y\rangle$ is given by a Kronecker delta, $\langle x|y\rangle = \delta_{xy}$.\\
\\
With linearity and complex conjugation, this defines an inner product between any two kets, $|f\rangle,|g\rangle \in \mathbb{H}^1(t)$. Note the overloading of notation such that $\langle f|g\rangle$ is both a statement and its truth value. Thus, $\mathbb{H}^1(t)$ is a Hilbert space, the basic conditional clauses of rule I are an orthonormal basis, and the space of bras is the dual space.\\
\\
\textbf{Definition:} The \textbf{position function} of the ket $|f\rangle \in \mathbb{H}^1(t)$ is the mapping, $\mathrm{D}\rightarrow\mathbb{C}$, $\forall x \in \mathrm{D}, x \rightarrow \langle x | f \rangle $.\\
\\
Later the position function will be identified with the restriction of the wave function to $\mathrm{D}$. It is here termed ``position function'' because it is discrete and because a wave equation is not assumed.

In this formal language, relative magnitudes are important in weighted logical \textsc{or}, but absolute magnitude has no meaning. It is easy in common language to construct phrases containing redundant words. \textit{``The black piece of coal''} is not the same phrase as \textit{``the piece of coal''}, but both have the same meaning. Similarly, for any complex number $a$, the clause $|f\rangle$ means exactly the same thing as $a|f\rangle$. When not part of a larger construction containing $+$, $a$ has the role of a redundant word.

The resolution of unity is found by expanding a ket in a normalised basis
\begin{equation}\label{Eq:2.5.1}
\ | f \rangle = \sum_{x \in \mathrm{D}} | x \rangle \langle x | f \rangle.
\end{equation}
Hence
\begin{equation}\label{Eq:2.5.2}
\ 1 = \sum_{x \in \mathrm{D}} | x \rangle \langle x | .
\end{equation}

The inner product is strictly a finite sum with $N$ terms, where $N = 8\nu^3$ is large. The formal limit $N\rightarrow \infty$, $\chi\rightarrow 0$ is strictly not part of the model, and is only to be taken at the final stage of calculation. With this in mind, it is convenient to normalize basis kets,
\begin{equation}\label{Eq:2.5.3}
\forall x,y \in \mathrm{D},  \langle x | y \rangle = \chi^{-3} \delta_{xy}.
\end{equation}
With this normalisation, the resolution of unity takes the form:
\begin{equation}\label{Eq:2.5.4}
\ 1 = \chi^3 \sum_{x \in \mathrm{D}} | x \rangle \langle x | .
\end{equation}

\subsection{Multiparticle kets}\label{sec:2.6}
\label{Multiparticle kets}
\begin{flushleft}
\textbf{RULE Va.} $|\rangle$ is the formal conditional clause, \textit{``If the first measurement at time $ t $ were to find no particle, \ldots''}.
\end{flushleft}
\textbf{RULE Vb.} $\langle |$ is the formal consequential clause, \textit{``\ldots, then a second measurement at time t would find no particle''}.\\
\\
\textbf{Definition:} Let $\mathbb{H}^0$ be the one dimensional space spanned by $|\rangle$.\\
\\
\textbf{Postulate:} The space of kets for $n$ particles of the same type is given by the $n^{th}$ tensor power $\mathbb{H}^n \equiv (\mathbb{H}^1)^{\otimes n} \equiv \underbrace{\mathbb{H}^1\otimes\cdots\otimes \mathbb{H}^1}_{n}$\\
\\
\textbf{RULE VIa.} $|x_1\rangle |x_2\rangle \ldots |x_n\rangle$ is the formal conditional clause, \textit{``If, for each of $ n $ particles, the measured position at time $ t $ of the $ i^{th} $ particle were $ x_i $, \ldots''}.\\
\\
\textbf{RULE VIb.} $\langle x_1|\langle x_2|\ldots\langle x_1|$ is the formal consequential clause, \textit{``\ldots, then, for each of $ n $ particles in a second measurement at time $ t $, the measured position of the $ i^{th} $ particle would be $ x_i $''}.\\
\\
\textbf{Postulate:} The space of any number of particles of the same type, $\gamma$, is $\mathbb{H}_\gamma \equiv \bigoplus\limits_{n} \mathbb{H}^n$\\
\\
The direct sum allows statements about an uncertain number of particles, using weighted logical \textsc{or}, \textit{``If, for each of $ n $ or $ m $ particles, but more likely $ n $ than $ m $, \ldots''}, etc. Since an $n$ particle ket cannot be an $m$ particle ket, the braket between kets of different numbers of particles is zero. For $|f\rangle=|f_1\rangle\ldots|f_n\rangle \in \mathbb{H}^n$, $|g\rangle=|g_1\rangle\ldots|g_n\rangle \in \mathbb{H}^n$,
\begin{equation}\label{Eq:2.6.2}
\langle f | g \rangle = \prod_{i=1}^{n} \langle f_i | g_i \rangle ,
\end{equation}
as is required for independent particles by the probability interpretation (section \ref{sec:2.10}). \\
\\
\textbf{Postulate:} The space of particles is $\mathbb{H} \equiv \bigoplus\limits_{\gamma} \mathbb{H}_{\gamma}$.\\
\\
\textbf{RULE VIIa.} $|x_1;x_2;\ldots;x_n \rangle$ is the formal conditional clause \textit{``If, for $ n $ identical particles, measured positions at time $ t $ were $ x_1, x_2, \ldots, x_n $''}.\\
\\
\textbf{RULE VIIb.} $\langle x_1;x_2;\ldots;x_n |$ is the formal consequential clause \textit{``then, for $ n $ identical particles, measured positions at time $ t $ would be $ x_1, x_2, \ldots, x_n $''}.\\
\\
\textbf{Postulate:} Since switching identical particles makes no difference to the physical situation, multiparticle space is \textbf{Fock space}, $\mathbb{F} \equiv \bigoplus\limits_{n} S\mathbb{H}^n$ where $S$ means that groups of tensor indices referring to the same type of particle are symmetrised for Bosons and antisymmetrised for Fermions.

\subsection{Momentum space}\label{sec:2.7}
\begin{flushleft}
\textbf{Definition:} For a 3-vector, $p$, at the origin, define the \textbf{momentum ket}, $|p\rangle$, as a sum of position kets:
\begin{equation}\label{Eq:2.7.1}
| p \rangle = \left(\tfrac{1}{2\pi}\right)^{3/2}\chi^3\sum_{x\in\mathrm{D}}e^{ix\cdot p}|x\rangle,
\end{equation}
where the dot product uses the Euclidean metric.
\end{flushleft}
The Euclidean metric in \eqref{Eq:2.7.1} has no direct bearing on a physical metric, and merely defines momentum kets as linear combinations of basic conditional clauses. The inner product with $|x\rangle$ defines a \textbf{plane wave},
\begin{equation}\label{Eq:2.7.2}
\langle x | p \rangle = \left(\tfrac{1}{2\pi}\right)^{3/2}e^{ix\cdot p}.
\end{equation}
\textbf{Definition:} $|p\rangle$ is a \textbf{plane wave ket} with \textbf{momentum} $p$. \\
\\
This is the fundamental definition of 3-momentum in this approach. It is justified because it will be found that $p$ is a conserved quantity which corresponds precisely to the classical notion of momentum.\\
\\
\textbf{Definition:} \textbf{Momentum space} is the 3-torus, $\mathrm{M} \equiv (-\frac{\pi}{\chi},\frac{\pi}{\chi}]^3 \subset \mathbb{R}^3$.\\
\\
There are momentum kets $|p\rangle$ in $\mathbb{H}^1$ for continuum values of $p\in\mathrm{M}$ (since they're just linear combinations of basis kets $|x\rangle$), but a discrete subset of momentum kets,
\begin{equation}\label{Eq:2.7.3}
\left\{|p\rangle, p\in \mathrm{M}_{\mathrm{D}} = \mathrm{M} \cap (\chi_p \mathbb{Z})^3\right\},
\end{equation}
is a basis for $\mathbb{H}^1$, where lattice spacing for $\mathrm{M_D}$ is given by $\chi_p = \pi / (\chi\nu)$. Using discrete transforms, Fourier inversion is exact. The resolution of unity in momentum space is
\begin{equation}\label{Eq:2.7.4}
\chi_p^3 \sum_{p\in \mathrm{M_D}} |p\rangle \langle p | = 1.
\end{equation}\\
\\
\textbf{Definition:} For $|f\rangle \in \mathbb{H}^1(t)$, determined by measurement at time $x^0 = t$ in discrete coordinates, $\mathrm{D}$, the \textbf{momentum space wave function} $F:\mathrm{M} \rightarrow \mathbb{C}$ is $p \rightarrow F(p) = \langle p | f \rangle$.\\
\\
In particular, for the position ket $|z\rangle$, the momentum space wave function is, for $p \in \mathrm{M}$,
\begin{equation}\label{Eq:2.7.5}
p \rightarrow \langle p | z \rangle = \left(\tfrac{1}{2\pi}\right)^{3/2}e^{-iz\cdot p}.
\end{equation}
It is straightforward to show that, for $x,y \in \mathrm{D}$,
\begin{equation}\label{Eq:2.7.6}
\int_{\mathrm{M}}d^3p\, \langle x|p \rangle\langle p|y \rangle = \left(\tfrac{1}{2\pi}\right)^{3}\int_{\mathrm{M}}d^3p\, e^{-iy\cdot p}e^{ix\cdot p} = \chi^{-3} \delta_{xy} = \langle x|y \rangle.
\end{equation}
Thus, Fourier inversion holds using the integral on momentum space; for any $|f\rangle \in \mathbb{H}^1(t)$,
\begin{equation}\label{Eq:2.7.7}
\int_{\mathrm{M}}d^3p\, \langle x|p \rangle\langle p|f \rangle =  \int_{\mathrm{M}}d^3p\, \chi^3 \sum_{y \in \mathrm{D}} \langle x|p \rangle\langle p|y \rangle \langle y| f \rangle = \langle x| f \rangle.
\end{equation}
So, we can identify the sum over discrete momenta with an integral over $\mathrm{M}$,
\begin{equation}\label{Eq:2.7.8}
1 \equiv \chi_p^3\sum_{p \in \mathrm{M_D}} |p \rangle\langle p| \equiv \int_{\mathrm{M}} d^3p\, |p \rangle\langle p|.
\end{equation}
Then for any $|f\rangle \in \mathbb{H}^1(t)$, $q \in \mathrm{M}$
\begin{equation}\label{Eq:2.7.9}
\langle q|f\rangle \equiv \chi_p^3\sum_{p \in \mathrm{M_D}} \langle q|p \rangle\langle p|f\rangle \equiv \int_{\mathrm{M}} d^3p\, \langle q|p \rangle\langle p|f\rangle.
\end{equation}
Thus, for any $p,q \in \mathrm{M}$, $\langle q|p\rangle = \delta(p-q)$. It is perhaps unexpected that the Dirac delta function on the test space of momentum space wave functions has an exact representation as a smooth function,
\begin{equation}\label{Eq:2.7.10}
\delta(p-q) \equiv \left(\tfrac{1}{2\pi}\right)^{3} \chi^3 \sum_{x \in \mathrm{D}}e^{ix\cdot (p-q)}.
\end{equation}

\subsection{Smooth representation}\label{sec:2.8}
\begin{flushleft}
\textbf{Definition:} $\mathrm{D}$ is embedded into the \textbf{continuum coordinate system}, $\mathrm{C}$,
\begin{equation}\label{Eq:2.8.1}
\mathrm{D} \subset \mathrm{C} \equiv (-\chi\nu, \chi\nu]^3 \subset \mathbb{R}^3.
\end{equation}
\end{flushleft}
\textbf{Definition:} For any $x \in \mathrm{C}$ we may define the \textbf{position ket} 
\begin{equation}\label{Eq:2.8.2}
|x\rangle = \chi_p^3 \sum_{p \in \mathrm{M_D}} |p\rangle \langle p | x \rangle = \int_{\mathrm{M}} d^3p\, |p \rangle \langle p | x \rangle.
\end{equation}
\textbf{Definition:} The \textbf{wave function} for $|f(t)\rangle \in \mathbb{H}^1(t)$ is $f(t): \mathrm{C} \rightarrow \mathbb{C}$ with
\begin{equation}\label{Eq:2.8.3}
x \rightarrow f(t,x) = \langle x|f(t) \rangle = \chi^3 \sum_{z \in \mathrm{D}} \langle x|z \rangle \langle z | f(t) \rangle
\end{equation}
Expanding the wave function in momentum space gives, for $x\in \mathrm{C}$, 
\begin{equation}\label{Eq:2.8.4}
f(x)=\langle z|f\rangle = \int_{\mathrm{M}}d^3p\, \langle x|p \rangle \langle p | f \rangle = \left(\tfrac{1}{2\pi}\right)^{3/2} \int_{\mathrm{M}}d^3p\, e^{ix\cdot p}\langle p | f \rangle .
\end{equation}Wave functions are differentiable. The wave function for $|z\rangle$, $z\in \mathrm{C}$, is, for $x\in \mathrm{C}$,
\begin{equation}\label{Eq:2.8.5}
x \rightarrow f_z(x) = \int_{\mathrm{M}}d^3p\, \langle x|p \rangle \langle p | z \rangle = \left(\tfrac{1}{2\pi}\right)^3 \int_{\mathrm{M}}d^3p\, e^{i(x-z)\cdot p}
\end{equation}
It is easily verified that for $x,z \in \mathrm{D}$ $f_z(x)=\chi^{-3}\delta_{xz} = \langle x | z \rangle$. So, the position function is the restriction of the wave function to $\mathrm{D}$, and, for $z \in \mathrm{D}$, there is a one-one correspondence between the wave functions, $f_z(x)$, and basis kets, $|z\rangle$, such that smooth wave functions are a representation of a finite dimensional Hilbert space. For $p, q \in \mathrm{M}$
\begin{equation}\label{Eq:2.8.6}
\int_{\mathrm{C}} d^3x\, \langle p|x \rangle \langle x | q \rangle = \left(\tfrac{1}{2\pi}\right)^{3}\int_{\mathrm{C}} d^3x\, e^{-ix\cdot (p-q)} = \chi_p^{-3} \delta_{pq} = \langle p|q \rangle.
\end{equation}
So, by linearity, we can identify the sum over discrete coordinates with an integral. The identity operator $1:\mathbb{H}^1 \rightarrow \mathbb{H}^1$ can be written
\begin{equation}\label{Eq:2.8.7}
1 \equiv \chi^3 \sum_{x \in \mathrm{D}} |x \rangle \langle x | \equiv \int_{\mathrm{C}} d^3x\, |x \rangle \langle x |.
\end{equation}
Then for any $|f\rangle \in \mathbb{H}^1$, $y \in \mathrm{C}$
\begin{equation}\label{Eq:2.8.8}
\langle y|f \rangle = \chi^3 \sum_{x \in \mathrm{D}} \langle y |x \rangle \langle x |f\rangle = \int_{\mathrm{C}} d^3x\, \langle y |x \rangle \langle x |f \rangle.
\end{equation}
and for any $x,y \in \mathrm{C}$ $\langle x|y \rangle = \delta(x-y)$ where the Dirac delta is a smooth function:
\begin{equation}\label{Eq:2.8.9}
\delta(x-y) \equiv (\tfrac{\chi_{p}}{2\pi})^3 \sum_{p \in \mathrm{M_D}} e^{i(x-y)\cdot p} \equiv \int_{\mathrm{M}} d^3p\,e^{i(x-y)\cdot p} .
\end{equation}

\subsection{Bounds}\label{sec:2.9}
Momentum space is the 3-torus $\mathrm{M}$, which is not covariant. The theory would break down if physical momentum could exceed $p_{\mathrm{max}}=\pi/\chi$, where $\chi$ is the lower bound of small lattice spacing, not the spacing appropriate to a given apparatus. In conventional units the components of momentum have a theoretical bound $p_{\mathrm{max}}=\pi\hbar c/\chi$. If Planck length is the smallest unit inherent in nature, the theoretical bound on the energy of an electron is $3.8 \times 10^{28}$eV, well beyond any reasonable level. Thus, in practice, physical momentum does not approach the bound and there is not an issue. 

In fact, there is a much lower bound on energy-momentum since an interaction between a sufficiently high energy electron and any electromagnetic field leads to pair creation (the Greisen-Zatsepin-Kuz'min limit on the energy of cosmic rays is $5 \times 10^{19}$eV \cite{Greisen}
\cite{Zatsepin}
). It follows from conservation of energy that the total energy of a system is bounded provided that energy has been bounded at some time in the past. This is true whenever an energy value is known since a measurement of energy creates an eigenket with a definite value of energy. Then momentum is also bounded, by the mass shell condition. The probability of finding a momentum above the bound is zero, and we assume that, for physically realizable states, $\langle p|f \rangle$ vanishes above the bound on each component of momentum. The bound depends on the system under consideration, but without needing to specify a least bound, we may reasonably assume that momentum is always much less than $\pi /(4\chi)$.

A theoretical bound on momentum might introduce a problem of principle for Lorentz transformation. If a high energy electron were boosted beyond the bound it might appear after the boost with a low energy, or with opposite direction of momentum. However, realistic Lorentz transformation means that macroscopic matter (i.e. the reference frame) is physically boosted by the amount of the transformation. In practice, Lorentz transformation cannot boost momentum beyond the level for which it is consistently defined.

The non-physical periodic property of $\langle p|f \rangle$  can removed by the substitution $\Theta_{\mathrm{M}}(p)\langle p|f \rangle \rightarrow \langle p|f \rangle$, where $\Theta_{\mathrm{M}}(p)=1$ if $p\in\mathrm{M}$ and $\Theta_{\mathrm{M}}(p)=0$ otherwise. With the replacement of the Euclidean dot product with Minkowski dot product (which takes place naturally in the solution of the Dirac equation, appendix \ref{Ap:E}), the expansion of the wave function in momentum space \eqref{Eq:2.8.4} is identical to the standard form in relativistic quantum mechanics, up to normalisation, and can be put into a manifestly covariant form:
\begin{equation}\label{Eq:2.9.1}
\begin{split}
f(x) &= \left(\tfrac{1}{2\pi}\right)^{3/2} \int_{\mathbb{R}^3}d^3p\, \langle p|f \rangle e^{-ix\cdot p}\\
&= \left(\tfrac{1}{2\pi}\right)^{3/2} \int_{\mathbb{R}^3}\frac{d^3p}{2p^0} F(p) e^{-ix\cdot p}\quad  \text{where}\quad  F(p)=2p^0\langle p|f \rangle \\
&= \left(\tfrac{1}{2\pi}\right)^{3/2} \int_{\mathbb{R}^4}d^4p F(p) e^{-ix\cdot p}\delta(p^2-m^2).
\end{split}
\end{equation}

\subsection{Probability interpretation}\label{sec:2.10}
To make the formal language precise, we must assign numerical values to the complex numbers introduced in rule II, i.e. we must determine magnitude and phase. Phase contains information on the evolution of kets, and will be considered later. Magnitude will be determined from probability. It only makes sense to talk about probability when we are actually going to do a measurement. When we are actually going to do the measurement, a statement about hypothetical measurement, in the subjunctive mood, automatically becomes a statement about real measurement, in the future tense. This being the case, truth values for hypothetical results must be replaced by truth values for future events, i.e. probabilities, when experiments are actually done.

In a typical measurement in quantum mechanics we study a particle in near isolation. The suggestion is that there are too few ontological relationships to create the property of position and that measurement introduces interactions which generate position. In this case, prior to measurement, position does not exist and the state of the system is not labelled by a position ket. Instead, Hilbert space is used to provide a label containing information about the about the probability of what would happen in measurement. To associate a ket, $|f\rangle$, with a particular physical state it is necessary and sufficient to specify the magnitude and phase of $\langle x|f\rangle$ from empirical data. If we set up many repetitions of a system described by the initial measurement results, $f$, and record the frequency of each result, $x$, then for a large number of repetitions the relative frequency of $x$ tends to the probability, $P(x|f)$, of finding the particle at $x$. Thus, in the first instance, amplitudes of the components $\langle x|f\rangle$ are determined from the probabilities of measurement results, not the other way about. In practice, they are determined from the results of previous measurements for which the results are known, together with the Schr\"{o}dinger equation (section \ref{sec:3.1}).\\
\\
\textbf{Postulate:} For the ket $|f\rangle \in \mathbb{H}^1(t)$, the magnitudes of the coefficients, $\langle x|f\rangle$ are defined such that
\begin{equation}\label{Eq:2.10.1}
\frac{|\langle x|f\rangle |^2}{\langle f|f\rangle} = P(x|f).
\end{equation}
\textbf{Definition:} If $\langle f|f\rangle =1$ then $ |f\rangle $ is said to be \textbf{normalised}.

\subsection{Observables}\label{sec:2.11}

Since only a general principle has been used that it is possible to measure position, it is necessary to discuss other observables. The question as to what other observables exist cannot be discussed until after a treatment of interactions between particles. It will be assumed that all observables are a product of physical laws arising from particle interactions. A full analysis of a given measurement would require that the measurement apparatus as well as the system being measured be treated as a multiparticle system in Fock space, in which time evolution for the interacting theory is known. Here general considerations are discussed on the assumption that interactions will be described by linear maps on Fock space and that measurement is always a physical process describable in principle as a combination of interaction operators. For qed this will mean that all observables depend only on the electric current operator and the photon field operator. A complete resolution of the measurement problem would demonstrate the projection postulate for any given apparatus and has not been given. The argument given below makes the projection postulate reasonable by reducing all measurement to measurement of position. The view is that if we find a physical process satisfying the projection postulate then we may say it defines an observable quantity.

Measurement has two effects on the state of a particle, altering it due to the interaction of the apparatus with the particle, and also changing the information we have about the state. New information causes a change of ket, even in the absence of physical change because the ket is just a label for available information. Then the collapse of the wave function is in part the effect of the apparatus on the particle, and in part the effect on conditional probability when the condition becomes known. This inverts the measurement problem; collapse represents a change in information due to a new measurement but Schr\"{o}dinger's equation requires explanation --- interference patterns are real. The requirement for a wave equation will be found in \ref{sec:3.1}.

Classical probability theory describes situations in which every parameter exists, but some are not known. Probabilistic results come from different values taken by unknown parameters. We have a similar situation here, but now the unknowns are not describable as parameters. We assume no relationships between particles bar those generated by physical interaction. An experiment is described as a large configuration of particles incorporating the measuring apparatus as well as the process being measured. The configuration has been partially determined by setting up the experimental apparatus, reducing the possibilities to those with definite outcomes to the measurement. It is impossible, even in principle, to determine every detail of the configuration since the determination of each detail requires measurement, which in turn requires a larger apparatus containing new unknowns in the configuration of particles. Thus there is always a lack of determination of initial conditions leading to randomness in the outcome, whether or not there is a fundamental indeterminism in nature. 

When we do a measurement, $K$, we get a definite result, a terminating decimal or $n$-tuple of terminating decimals read off the measurement apparatus. Let the possible results be $k_i \in \mathbb{Q}^n$ for $i=1,\ldots,m$ . We assume that the dimension of $\mathbb{H}^1$ is greater than $m$; this must be so if all measurements are reducible to measurements of position, and can be ensured by the choice of a lattice finer than the resolution of measurement. Each physical state is associated with a ket, labelled by the measurement result, so that if the measured result is $k_i$ then the ket is $|k_i \rangle$. The empirical determination of $|k_i \rangle$ as a member of $\mathbb{H}^1$ requires that we draw from experimental data the value of the inner product $\langle k_i | f \rangle$ for an arbitrary ket, $|f \rangle$. Without loss of generality $|k_i \rangle$ and $|f \rangle$ are normalised. By assumption, measurement of $K$ is reducible to a set of measurements of position, so that each $k_i$ is in one to one correspondence with the positions $y_i$ of one or more particles used for the measurement (e.g. $y_i$ may be the positions of one or more pointers). Then, 
\begin{equation}\label{Eq:2.11.1}
|\langle k_i|f\rangle |^2 = |\langle y_i|f\rangle |^2 = P(y_i|f) = P(k_i|f)
\end{equation}
is the probability that a measurement of $K$ has result $k_i$, given the initial ket $|f\rangle \in \mathbb{H}^1$. It follows from $\langle x|y \rangle = \delta_{xy}$ that $\langle k_i|k_j \rangle = \delta_{ij} = \langle y_i|y_j \rangle$. So, if the result is $k_i$ it is definitely $k_i$ and cannot at the same time be $k_j$ with $i\neq j$.

Measurement with result, $k_i$, implies a physical action on a system and is represented by the action of an operator, $K_i$, on Hilbert space. If a quantity is measurable we require that there is an element of physical reality associated with its measurement, by which we mean that the configuration of particles necessarily becomes such that the quantity has a well defined value. In practice this means that, in the limit in which the time between two measurements goes to zero, a second measurement of the quantity necessarily gives the same result as the first. It follows that $K_i$ is a projection operator (the projection postulate), 
\begin{equation}\label{Eq:2.11.2}
K_i = | k_i \rangle \langle k_i |
\end{equation}
The projection postulate is too restrictive to describe all numerical quantities used in the classical description of nature, and will be relaxed after a discussion of expectations (section \ref{sec:2.13}).

The expectation of the result from a measurement of $K$, given the initial normalised ket, $|f\rangle \in \mathbb{H}^1$, is
\begin{equation}\label{Eq:2.11.3}
\langle K \rangle \equiv \sum_i k_i P(k_i| f) = \sum_i \langle f|k_i \rangle k_i \langle k_i | f \rangle = \langle f|K|f \rangle
\end{equation}
\textbf{Postulate:} The Hermitian operator, $K = \sum_i |k_i \rangle k_i \langle k_i | $, is called an \textbf{observable}. $k_i$ is the \textbf{value} of $K$ in the ket $|k_i\rangle$.

Using \eqref{Eq:2.11.1} the probability that operators describing the interactions comprising the measurement of $K$ combine to give the result $K_i$ is
\begin{equation}\label{Eq:2.11.4}
P(k_i|f)=|\langle k_i | f \rangle |^2 = \langle f | k_i \rangle \langle k_i | f \rangle = \langle f | K | f \rangle .
\end{equation}
Then $P(k_i|f)$ can be understood as a classical probability function, where the random variable runs over the set of projection operators, $K_i$, corresponding to the outcomes of the measurement. The physical interpretation is that each $K_i$ represents a set of unknown configurations of particle interactions in measurement, namely that set of configurations leading to the result $k_i$. 

\subsection{The canonical commutation relation}\label{sec:2.12}
\begin{flushleft}
\textbf{Definition:} The \textbf{momentum operator}, $P^a=-i\partial^a:\mathbb{H}^1 \rightarrow \mathbb{H}^1$, is, for $a=1,2,3,$  
\begin{equation}\label{Eq:2.12.1}
P^a:|f\rangle\rightarrow -\int_{\mathrm{C}}d^3x\,|x\rangle i\partial^a \langle x|f \rangle
\end{equation}
\end{flushleft}Clearly $P^a$ is Hermitian and
\begin{equation}\label{Eq:2.12.2}
P^a|f\rangle = -\int_{\mathrm{C}}d^3x\,|x\rangle i\partial^a \chi_p^3 \sum_{p \in \mathrm{M_D}} \langle x|p \rangle \langle p|f \rangle = \chi_p^3 \sum_{p \in \mathrm{M_D}} |p \rangle p^a \langle p|f \rangle .
\end{equation}Similarly,
\begin{equation}\label{Eq:2.12.3}
P^a|f\rangle = \int_{\mathrm{M}}d^3p\, |p \rangle p^a \langle p|f \rangle.
\end{equation}
\textbf{Definition:} The \textbf{position operator}, $X^a:\mathbb{H}^1 \rightarrow \mathbb{H}^1$, is, for $a=1,2,3$
\begin{equation}\label{Eq:2.12.4}
X^a |f\rangle = \chi^3 \sum_{x \in \mathrm{D}} |x \rangle x^a \langle x|f \rangle
\end{equation}

From the property that the trace of a commutator in finite dimensional Hilbert space vanishes, $ \mathrm{Tr}([X^a,P^b]) = 0$, it follows that $[X^a,P^b]\neq i \delta_{ab}$, and the canonical commutation relation does not hold. If we formally define $\tilde{X}$ by
\begin{equation}\label{Eq:2.12.5}
\tilde{X}^a|f\rangle = \int_{\mathrm{C}}d^3x\, |x \rangle x^a \langle x|f \rangle.
\end{equation}
Then,
\begin{equation}\label{Eq:2.12.6}
P^b\tilde{X}^a|f\rangle = \int_{\mathrm{C}}d^3x\, |x \rangle i\delta_{ab} \langle x|f \rangle - \int_{\mathrm{C}}d^3x\, |x \rangle x^ai\partial^b \langle x|f \rangle = -i\delta_{ab} - \tilde{X}^aP^b|f\rangle.
\end{equation}
So,
\begin{equation}\label{Eq:2.12.7}
[\tilde{X}^a,P^b]= i \delta_{ab}.
\end{equation}
and we conclude that $X^a \neq \tilde{X}^a$ and that $\tilde{X}^a|f\rangle \notin \mathbb{H}^1$. 

\subsection{Classical correspondence}\label{sec:2.13}
In the classical correspondence we study the behaviour of systems containing a large number, $N$, of quantum motions (this is sometimes called the thermodynamic limit). A classical property is the expectation, \eqref{Eq:2.11.3}, of the corresponding observable in the limit $N \rightarrow \infty$ (not $\hbar \rightarrow 0$ as sometimes stated; Planck's constant is simply a change of scale from natural to conventional units and it would be meaningless to let it go to zero). For example, the centre of gravity of a macroscopic body is a weighted average of the positions of the elementary particles which constitute it. Schr\"{o}dinger's cat is definitely either alive or dead because, consisting as it does of a large number of elementary particles, its properties are expectations obeying classical laws derived from \eqref{Eq:2.11.3}, but the ket simply encodes probability and the cat may be described as a superposition until the box is opened.\\
\\
\textbf{Postulate: }A \textbf{measurement} of a physical quantity is any physical process such that a determination of the quantity is possible in principle. \\

In keeping with the considerations of section \ref{sec:2.11}, we assume that the existence of a value for an observable quantity depends only on the configuration of matter. If a configuration of matter corresponds to an eigenket of an observable operator then the value of that observable exists independently of observation and is given by the corresponding eigenvalue. In classical physics there is sufficient information to determine the motion at each instant between the initial and final ket, up to experimental accuracy. Intermediate kets are similarly determinate and may be calculated in principal by the processing of data already gathered, or which could be gathered without physically affecting the measurement. So in classical physics intermediate states may be regarded as measured states, and we may say that they are \textbf{effectively measured}.

The projection postulate is required if the results of measurement are to be used to name states in Hilbert space, but classical quantities can also be defined from Hermitian operators when this is not the case. To say that a Hermitian operator has a well defined value in a given state, a measurement should necessarily yield that value as the expectation of the operator:\\
\\
\textbf{Postulate:} For kets consisting of large numbers of particles, the \textbf{classical value} of an observable quantity is given by the expectation of the corresponding Hermitian operator (irrespective of whether the ket is an eigenket).\\
\\
This is weaker than the projection postulate, which requires an eigenket (in which the value is trivially given by the expectation). The reason is that it will be found in section \ref{sec:6} that the classical electromagnetic field is given by the expectation of the photon field operator.

\section{Particles}\label{sec:3}
\subsection{The Schr\"{o}dinger equation}\label{sec:3.1}
The inner product allows us to calculate probabilities for the outcome of a measurement provided that we know the ket describing hypothetical measurement at the time of measurement. This is only useful if we can calculate the ket at any time, $t$, from a known previous measurement result. Hilbert space refers to measurement at time, $t$, so that $|f(t)\rangle \in \mathbb{H}(t)$, where $t$ is a parameter and we isomorphically identify $\mathbb{H}(t) = \mathbb{H}$ for all $t$. The position ket $|x\rangle$ at time $x^0 = t$ will be denoted by $|t,x\rangle$. Since $\mathbb{H}$ has a finite basis, it is required to review the arguments for the Schr\"{o}dinger equation.\\
\\
\textbf{Postulate:} If at time $t_0$ the ket is $|f(t_0)\rangle$, then the ket at time $t$ is given by the \textbf{time evolution operator}, $U(t, t_0): \mathbb{H} \rightarrow \mathbb{H}$, such that $|f(t)\rangle = U(t,t_0)|f(t_0)\rangle$.\\
\\
If the ket at time $t_0$ was either $|f(t_0)\rangle$ or $|g(t_0)\rangle$, then it will evolve into either $|f(t)\rangle$ or $|g(t)\rangle$ at time $t$. Any weighting in \textsc{or} will be preserved. So, $ U $ is linear 
\begin{equation}\label{Eq:3.1.1}
U(t,t_0)(a|f(t_0)\rangle + b|g(t_0)\rangle) = aU(t,t_0)|f(t_0)\rangle + bU(t,t_0)|g(t_0)\rangle.
\end{equation}

Irrespective of whether a model of discrete particles might appear continuous on the large scale, the evolution of kets is expected to be continuous because kets are not physical states of matter, but are rather probabilistic statements about what might happen in measurement, given current information. Probabilities describe our ideas concerning the likelihood of events. Whether or not reality is fundamentally discrete, probability is properly described on a mathematical continuum. A discrete interaction will not lead to a discrete change in probability because we do not have exact information on when the interaction takes place. This being so, time evolution will be modelled by a continuous operator valued function of time, $U$. Since local laws of physics are always the same, and $U$ does not depend on the ket on which it acts, the form of the evolution operator for a time span $t$, $U(t)=U(t+t_0,t_0)$, does not depend on $t_0$. We require that the evolution in a span $t_1 + t_2$ is the same as the evolution in $t_1$ followed by the evolution in $t_2$, and is also equal to the evolution in $t_2$ followed by the evolution in $t_1$, $U(t_2)U(t_1) = U(t_2+t_1) = U(t_1)U(t_2)$. In zero time span, there is no evolution. So, $U(0)$ does not change the ket; $U(0)=1$. Using negative $t$ reverses time evolution (put $t = t_1 = -t_2$); $U(-t)=U(t)^{-1}$.

Since kets can be chosen to be normalised we may require that $U$ conserves the norm, i.e. for all $|g\rangle$, $\langle g|U^{\dagger} U|g\rangle = |U|g\rangle |^2 = ||g\rangle |^2 = \langle g|g \rangle$. This is sufficient to show that $U$ is unitary (appendix A). Thus the conditions of Stone's theorem \cite{Stone} (appendix B) are satisfied and we have that there exists a Hermitian operator $H$, the Hamiltonian, such that $\dot{U}(t)=-iHU(t)$. This has solution $U(t)=e^{-iHt}$. The Schr\"{o}dinger equation and Newton's first law ($H=E = \text{const}$) follow immediately. $E$ is identified with energy and $m$ with mass.

In a general problem in quantum theory, an initial condition is described by a ket $|f \rangle$ with momentum space wave function $\langle p|f \rangle$, and such that the discrete position function is uniquely embedded into the smooth wave function on $\mathbb{R}^3$, \eqref{Eq:2.8.4}. Solving the Schr\"{o}dinger equation extends the wave function to $\mathbb{R}^4$, \eqref{Eq:2.9.1}. Then the position function at any time, and in any discrete coordinate system is found restricting to discrete values. Thus we do not require the existence of a physical continuum to define quantum theory using smooth wave functions.

\subsection{Quantum covariance}\label{sec:3.2}
If time and position are not properties of prior space or spacetime, but only of relationships found in matter, then it follows that the fundamental properties of elementary particles have no dependency on time or position. This is expressed in the principle that, \textit{the fundamental behaviour of matter is always and everywhere the same}. Incorporated in this law is the notion that local, physically realised, coordinate systems may always be established by an observer in the same way. From this we may infer the general principle of relativity, \textit{local laws of physics are the same irrespective of the coordinate system which a particular observer uses to quantify them}. In classical physics, laws which are the same in all coordinate systems are most easily expressed in terms of invariants, known as tensors. Then the most directly applicable form of the principle of general relativity is the principle of general covariance, \textit{the equations of physics have tensorial form}.

General covariance applies to classical vector quantities under the assumption that they are unchanged by measurement. But in quantum mechanics measured values arise from the action of the apparatus on the quantum system, creating an eigenket of the corresponding observable operator and we cannot generally assume the existence of a tensor independent of measurement. In practice a change of reference frame necessitates a change of apparatus (either by accelerating the apparatus or by switching to a different apparatus). A lattice describes possible values taken from measurement by a particular apparatus. Eigenkets of displacement are determined by this lattice, i.e. by the properties and resolution of a particular measuring apparatus. So, in general, eigenkets in one frame are not simultaneously eigenkets of a corresponding observable in another frame using another apparatus (c.f. non-commutative geometry \cite{Connes}
). For the same reason classical tensor quantities do not, in general, correspond to tensor observables.

The broad meaning of covariance is that it refers to something which varies with something else, so as to preserve certain mathematical relations. If covariance is not now to be interpreted as manifest covariance or general covariance as applicable to the components of classical vectors, then a new form of covariance, \textbf{quantum covariance}, is required to express the principle of general relativity, that local laws of physics are the same in all reference frames. Quantum covariance will mean that local laws of physics have the same form in any reference frame but not that the same physical process may be described identically in different reference frames, since the reference frame, i.e. the choice of apparatus, can affect both the process under study and the description of that process. Since coordinates are determined by physical measurement which has finite resolution, under transformation of the coordinate system (passive Lorentz transformation) there is also a change of basis for Hilbert space. Quantum covariance observes that, since the choice of basis is arbitrary and observer dependent, and since Hilbert space contains a continuum of kets $|x\rangle$ for $x \in \mathbb{R}^3$, any breaking of manifest covariance by the choice of basis is irrelevant. \\
\\
\textbf{Postulate:} \textbf{Quantum covariance} will mean that the wave function, \eqref{Eq:2.9.1}, is defined on a continuum, while the inner product is discrete, and that, in a change of reference frame, the lattice and inner product appropriate to one reference frame are replaced with the lattice and inner product of another.\\

Thus, from an initial position function defined on $\mathrm{C}$, the position function at any time is given by
\begin{equation}\label{Eq:3.2.1}
\langle x|f \rangle = f(x)|_{\mathrm{S}},
\end{equation}
and if, in a change of reference frame, the spacetime coordinate system $\mathrm{S}$ is replaced by $\mathrm{S'}$, the new position function is given by
\begin{equation}\label{Eq:3.2.2}
\langle x|f \rangle = f(x)|_{\mathrm{S'}}.
\end{equation}
We have seen that the consistency of quantum covariance is ensured if the support of $\langle p|q \rangle$ is bounded as described in section \ref{sec:2.9}. 

The general form of a linear operator, $O$ on $\mathbb{H}$, is, for some complex valued function $O(x,y)$,
\begin{equation}\label{Eq:3.2.3}
O = \chi^3 \sum_{x,y \in \mathrm{D}} |x\rangle O(x,y) \langle y|.
\end{equation}
According to quantum covariance, this expression has an invariant form under a change of reference frame. This has important implications for the definition of quantum fields. The invariance of operators under rotations is perhaps at first a little surprising, particularly when one considers the presumed importance of manifest covariance in axiomatic quantum field theory. It may be clarified a little with a nautical analogy. On a boat the directions fore, aft, port and starboard are invariant because they are defined with respect to the boat. Similarly operators are necessarily defined with respect to chosen reference matter and have an invariant form with respect to reference matter.

\subsection{Dirac particles}\label{sec:3.4}
It has been seen that the probability interpretation requires a first order Schr\"{o}dinger equation (section \ref{sec:3.1} and appendix \ref{Ap:B}). There is no covariant first order equation for a spinless particle\footnote{This applies to fundamental particles but does not preclude scalar composite or scalar ghost particles.} and, following Dirac \cite{Dirac}
, a spin index is added to the ket. When there is no ambiguity spin is suppressed. Explicitly,
\begin{equation}\label{Eq:3.3.1}
|x\rangle=|x,\alpha\rangle=|x\rangle_{\alpha}.
\end{equation}
I will use the normalisation:
\begin{equation}\label{Eq:3.3.2}
\forall x,y \in \mathrm{D}, \langle x,\alpha | y, \beta \rangle = \langle x|y \rangle_{\alpha\beta} = \chi^{-3} \delta_{xy} \delta_{\alpha\beta}.
\end{equation}
The inner product is
\begin{equation}\label{Eq:3.3.3}
\langle g|f \rangle = \chi^3 \sum\limits_{\mathrm{D}} \langle g|x \rangle \langle x|f \rangle =  \chi^3 \sum\limits_{\mathrm{D}} g_{\alpha}(x)^{\dagger} f_{\alpha}(x) ,
\end{equation}
where the summation convention is used for spin indices. Position functions defined on the discrete spacetime coordinate system $\mathrm{S}$ are embedded into smooth wave functions. Wave functions now have a spin index, $ f_{\alpha}(x)|_{\mathrm{S}} = \langle x|f\rangle_{\alpha}$, and the first order equation required by Stone's theorem \cite{Stone} (
appendix \ref{Ap:B}) is the Dirac equation:
\begin{equation}\label{Eq:3.3.4}
i \partial \cdot \gamma f(x) = m f(x).
\end{equation}
Using bold font to denote 3-vectors, the solution to the Dirac equation is (appendix \ref{Ap:E})
\begin{equation}\label{Eq:3.3.5}
f_{\alpha}(x) = (\tfrac{1}{2\pi})^{3/2}\sum\limits_{r=1}^{2}\int_{\mathrm{M}}d^3\boldsymbol{p}\,F(\boldsymbol{p},r)u_{\alpha}(\boldsymbol{p},r) e^{-x\cdot p},
\end{equation}
where $ F(\boldsymbol{p},r) $ is the momentum space wave function given at $ x^0 = 0 $ by
\begin{equation}\label{Eq:3.3.6}
F(\boldsymbol{p},r) = \langle \boldsymbol{p},r| f \rangle= (\tfrac{1}{2\pi})^{3/2} \chi^3 \sum\limits_{\mathrm{D}} u(\boldsymbol{p},r) ^{\dagger} f(0,\boldsymbol{x}) e^{i\boldsymbol{x}\cdot\boldsymbol{p}},
\end{equation}
$ p $ satisfies the mass shell condition, and $ u $ is a Dirac spinor having the form, for $r = 1,2$,
\begin{equation}\label{Eq:3.3.7}
u(\boldsymbol{p},r)=\sqrt{\frac{p^0 + m}{2p^0}}\begin{bmatrix}\zeta(r)\\
\frac{\boldsymbol{\sigma} \cdot \boldsymbol{p}}{p^0 + m}\zeta(r)
\end{bmatrix},
\end{equation}
where $\boldsymbol{\sigma} = (\sigma_1,\sigma_2,\sigma_3)$ are the Pauli spin matrices and $ \zeta $ is a two-spinor normalised so that $ \zeta_{\alpha}(r)^{\dagger} \zeta_{\alpha}(s)= \delta_{rs} $, where the summation convention is used for repeated spin indices. In this normalisation $ u_{\alpha}(\boldsymbol{p},r)^{\dagger}u_{\alpha}(\boldsymbol{p},s) = \delta_{rs} $ (appendix \ref{Ap:F}). It is common to choose a relativistic normalisation by multiplying Dirac spinors by $ \sqrt{2p^0} $. Since we ultimately divide by normalisation to calculate probability, this makes no difference to predictions. The normalisation used here is consistent with the idea that probability is observer dependent, and leads to simpler formulae.

\subsection{Antiparticles}\label{sec:3.5}
The treatment of the antiparticle modifies the St\"{u}ckelberg-Feynman \cite{St\"{u}ckelberg}\cite{Feynman1} 
interpretation by considering the mass shell condition. The Dirac equation is most readily understood as the equation of motion for a particle in its own proper time. If every particle has its own proper time, and if there is no other fundamental time, then it is natural to think that one particle's proper time can be reversed compared to that of another; antimatter is matter whose proper time is inverted compared to surrounding matter. A sign is lost in the mass shell condition, due to the squared terms, but a time-like vector with a negative time-like component provides a natural definition of negative mass, $m < 0$. So, permissible solutions of the Dirac equation, \eqref{Eq:3.3.4}, have positive energy, $ E=p^0>0 $, when $m$ is positive and negative energy when $m$ is negative. Complex conjugation reverses time, and the direction of momentum, while maintaining the probability interpretation and restores positive energy, and we also change the sign of mass, $ m\rightarrow -m $. Thus the negative energy solution is transformed and satisfies
\begin{equation}\label{Eq:3.4.1}
i\partial \cdot \bar{\gamma} f(x) = - mf(x),
\end{equation}
where $ \bar{\gamma} $ is the complex conjugate, $ \bar{\gamma}_{\alpha\beta}^j = \overline{\gamma_{\alpha\beta}^j} $. The solution is the wave function for the antiparticle
\begin{equation}\label{Eq:3.4.2}
f(x) = (\tfrac{1}{2\pi})^{3/2}\sum\limits_{r=1}^{2}\int_{\mathrm{M}}d^3\boldsymbol{p}\,F(\boldsymbol{p},r)\bar{v}(\boldsymbol{p},r) e^{-x\cdot p},
\end{equation}
where $F(\boldsymbol{p},r)$ is the momentum space wave function given by
\begin{equation}\label{Eq:3.4.3}
F(\boldsymbol{p},r) = (\tfrac{1}{2\pi})^{3/2} \chi^3 \sum\limits_{\mathrm{D}} \bar{v}(\boldsymbol{p},r)^{\dagger} f(0,\boldsymbol{x}) e^{i\boldsymbol{x}\cdot\boldsymbol{p}},
\end{equation}
$ p $ satisfies the mass shell condition, and $ \bar{v} $ is the complex conjugate of the Dirac spinor, for $r = 1,2$,
\begin{equation}\label{Eq:3.4.4}
v(\boldsymbol{p},r)=\sqrt{\frac{p^0 + m}{2p^0}}\begin{bmatrix}\frac{\boldsymbol{\sigma} \cdot \boldsymbol{p}}{p^0 + m}\zeta(r)\\
\zeta(r)
\end{bmatrix}.
\end{equation}
The spinor has the normalisation $ \bar{v}_{\alpha}(\boldsymbol{p},r)^{\dagger}\bar{v}_{\alpha}(\boldsymbol{p},s) = \delta_{rs} $. The use of $ \bar{v} $ in \eqref{Eq:3.4.2} and \eqref{Eq:3.4.3} will be reconciled with the more usual form when field operators are defined.

\subsection{Conserved current}\label{sec:3.6}
Since the interpretation is based on probability theory, we need a relativistic statement that probability is conserved. That is, we require a vector current density $ j^{a} $ such that 
\begin{equation}\label{Eq:3.5.1}
\partial_a j^a = 0,
\end{equation}
and the probability density for finding a particle at $x$ is
\begin{equation}\label{Eq:3.5.2}
j^0(x) = f(x)^{\dagger}f(x).
\end{equation}
\textbf{Definition:} The \textbf{Dirac adjoint} of a Dirac spinor, $ u $, is $ \hat{u}= u^{\dagger}\gamma^0 $.\\
\\
\textbf{Postulate:} \textbf{Current density} is $ j^a = \hat{f}\gamma^af $.\\
\\
With these definitions, current density satisfies,
\begin{equation}\label{Eq:3.5.3}
\partial_a j^a = \partial_a (f^{\dagger}\gamma^0\gamma^af) = (\partial_a f^{\dagger})\gamma^{a\dagger} \gamma^0 f + \hat{f}\gamma^a \partial_af = im \hat{f}f - im \hat{f}f =0,
\end{equation}
as is required of a conserved current.

\subsection{Creation operators}\label{sec:3.7}
In interactions, particles may be created and destroyed. The creation of a particle in an interaction is described by the action of a creation operator, and destruction is described by an annihilation operator. A change of state of a particle can be described as the annihilation of one state and the creation of another. Thus, a complete description of any process in interaction can be achieved through combinations of creation and annihilation operators. Creation and annihilation operators are linear operators, and incorporate the idea that when a particle is created it is impossible to distinguish it from any existing particle of the same type, so that they automatically (anti)symmetrise states of identical particles. A creation operator is closely associated with the state which it creates, and will be denoted as a ket, with an underline to distinguish it from a state.\\
\\
\textbf{Definition:} $\forall x \in \mathrm{D}$ the \textbf{creation operator} $ |\underline{x}\rangle: \mathbb{H} \rightarrow \mathbb{H} $  is $ \forall y, y^i \in \mathrm{D}, i=1, \ldots, n  $ 
:\begin{subequations}\label{Eq:3.6.1}
\begin{align}
&|\underline{x}\rangle: |\rangle \rightarrow|\underline{x}\rangle |\rangle = |x\rangle \\
&|\underline{x}\rangle: |y\rangle \rightarrow  |x;y\rangle = \tfrac{1}{\sqrt{2}} [|x\rangle |y\rangle + \kappa |y\rangle]|x\rangle]\\
&|\underline{x}\rangle: |y^1\rangle \ldots|y^n\rangle \rightarrow  \frac{1}{\sqrt{n+1}} \sum\limits_{i=0}^{n} \kappa^i |y^1\rangle \ldots|y^i\rangle |x\rangle |y^{i+1}\rangle \ldots|y^n\rangle
\end{align}
\end{subequations}
where $ |x\rangle $  appears in the $ i+1^{th} $ position in the $ i^{th} $ term of the sum. It is routine to show that $ \kappa=\pm 1 $  for Bosons and Fermions respectively. More generally, creation operators are defined by linearity:
\begin{equation}\label{Eq:3.6.2}
\forall |f\rangle \in \mathbb{H}^1, |\underline{f}\rangle = \chi^3 \sum\limits_{x \in \mathrm{D}} \langle x|f\rangle |\underline{x} \rangle.
\end{equation}
We have
\begin{equation}\label{Eq:3.6.3}
|\underline{x}\rangle: |y^1; \ldots;y^n\rangle \rightarrow  |x;y^1; \ldots;y^n\rangle.
\end{equation}
So the space of (anti)symmetric states $ \mathbb{F}\subset\mathbb{H} $ is generated from $\mathbb{H}^0=\{|\rangle\}  $ by creation operators. Physical states are elements of $ \mathbb{F} $.\\
\\
\textbf{Definition:} $ \forall |f\rangle \in \mathbb{H}^1 $, the \textbf{annihilation operator} $ \langle\underline{f}|:\mathbb{F}\rightarrow\mathbb{F} $ is the Hermitian conjugate of the creation operator $ |\underline{f}\rangle:\mathbb{F}\rightarrow\mathbb{F} $, $ \langle\underline{f}|= |\underline{f}\rangle^{\dagger} $ .

\section{Interactions}\label{sec:4}
\subsection{The interaction Hamiltonian}\label{sec:4.1}
By Stone's theorem \cite{Stone} (appendix \ref{Ap:B}) time evolution is modelled by a continuous operator, $ U(t):\mathbb{F}\rightarrow\mathbb{F} $, such that $U(t)=e^{-iHt}$. In a time interval, $t$, there either is, or is not, an interaction:\\
\\
\textbf{Postulate:} By the identification of the operations of vector space with weighted \textsc{or} between uncertain possibilities, time evolution including the possibility of interaction is described by a \textbf{Hamiltonian}, $ H:\mathbb{F}\rightarrow\mathbb{F} $, with 
\begin{equation}\label{Eq:4.1.1}
H=H_0+H_{\mathrm{int}},
\end{equation}
where $ H_{0} $ is the \textbf{free Hamiltonian}, and $ H_{\mathrm{int}} $ is a Hermitian operator describing an interaction between particles, called the \textbf{interaction Hamiltonian}. $ H_{\mathrm{int}} $ is defined with no component corresponding to the absence of interaction,
\begin{equation}\label{Eq:4.1.2}
\forall \boldsymbol{x}^i \in \mathrm{D}, \forall n\in \mathbb{N}, \langle \boldsymbol{x}^1;\ldots\boldsymbol{x}^n| H_{\mathrm{int}} |\boldsymbol{x}^1;\ldots\boldsymbol{x}^n\rangle = 0.
\end{equation}
In general, $ H_{\mathrm{int}} $ will be a sum of terms for different types of interaction. Here only one type of interaction will be considered. Thus, the evolution of a ket is given by
\begin{equation}\label{Eq:4.1.3}
\partial_0|f(t)\rangle=-iH|f(t)\rangle=-i(H_{0}+ H_{\mathrm{int}})|f(t)\rangle.
\end{equation}

It is convenient to separate interactions from free particle evolution by working in the interaction picture, so that
\begin{align}\label{Eq:4.1.4}
|f_{\mathrm{I}}(t)\rangle & = e^{iH_0t}|f(t)\rangle,\\
\label{Eq:4.1.5}
A_{\mathrm{I}}(t) & = e^{iH_0t}Ae^{-iH_0t},\\
\label{Eq:4.1.6}
H_{\mathrm{I}}(t) & = e^{iH_0t}H_{\mathrm{int}}e^{-iH_0t}.
\end{align}
As is common practice, the suffix, I, denoting the interaction picture will be dropped when there is no ambiguity. From (62), since $ |f(t)\rangle = e^{-iHt} |f(0)\rangle $,  evolution is given in the interaction picture by 
\begin{equation}\label{Eq:4.1.7}
U(t) = e^{iH_0t}e^{-iHt}
\end{equation}
Differentiating gives (appendix \ref{Ap:C}) 
\begin{equation}\label{Eq:4.1.8}
\dot{U}(t) = -iH_{\mathrm{I}}U(t),
\end{equation}
which has solution ($ U(0)=1 $)
\begin{equation}\label{Eq:4.1.9}
U(t) = e^{-iH_{\mathrm{I}}t}.
\end{equation}

\subsection{The Hamiltonian density}\label{sec:4.2}
We assume that we can define a Hermitian interaction density operator, $ I(x) $, having the same effect on a matter anywhere and at any time, as required by the general principle of relativity. By the identification of addition with logical disjunction, $ H_{\mathrm{I}} $ can be written as a sum:\\
\\
\textbf{Postulate:} The \textbf{Hamiltonian (or interaction) density}, $ I(x):\mathbb{F}\rightarrow\mathbb{F} $, is a Hermitian operator such that the interaction Hamiltonian is
\begin{equation}\label{Eq:4.2.1}
H_{\mathrm{I}}(x^0) = \chi^3 \sum\limits_{\mathrm{D}}I(x).
\end{equation}

\subsection{The perturbation expansion}\label{sec:4.3}
Without loss of generality, $ t_0=0 $. The discrete time interval is $ \mathrm{T}=\lbrace t_j=j\chi |j\in\mathbb{Z}, 0\le j \le n $ for some $n \in \mathbb{Z} \rbrace $. We have
\begin{equation}\label{Eq:4.3.1}
U(t_{j+1})=U(\chi)U(t_j)= (1-iH_{\mathrm{I}}(t_j)\chi)U(t_j).
\end{equation}
Iterating
\begin{align}\label{Eq:4.3.2}
U(t_{1})&= 1-iH_{\mathrm{I}}(t_0)\chi.\\
\label{Eq:4.3.3}
U(t_{2})&= (1-iH_{\mathrm{I}}(t_1)\chi)(1-iH_{\mathrm{I}}(t_0)\chi).
\end{align}
Expand after $n$ iterations
\begin{equation}\label{Eq:4.3.4}
U(t_{n})= 1-i\chi\sum\limits_{i=0}^{n-1}H_{\mathrm{I}}(t_i) + (-i\chi)^2\sum\limits_{j>i}\sum\limits_{i=0}^{n-1}H_{\mathrm{I}}(t_j)H_{\mathrm{I}}(t_i) + \ldots.
\end{equation}
Substituting \eqref{Eq:4.2.1}
\begin{equation}\label{Eq:4.3.5}
U(t_{n})= 1-i\chi^4\sum\limits_{i=0}^{n-1}I(x_i) + (-i\chi^4)^2\sum\limits_{j>i}\sum\limits_{i=0}^{n-1}I(x_j)I(x_i) + \ldots,
\end{equation}
where the sums are now over space as well as time. This can be rewritten
\begin{equation}\label{Eq:4.3.6}
U(t_{n})= 1+ \sum\limits_{k=1}^{n} (-i\chi^4)^k \sum\limits_{i_k>i_{k-1}} \ldots  \sum\limits_{i_2>i_1}  \sum\limits_{i_1=0}^{n-1}I(x_{i_k}) \ldots I(x_{i_2})I(x_{i_1}).
\end{equation}

\subsection{Time-ordered diagrams}\label{sec:4.4}
Any operator on Fock space, $ \mathbb{F} $, can be written as a sum of products of creation and annihilation operators. The change of state associated with an interaction can be described as the annihilation of one state and the creation of another. Thus, a complete description of any process in interaction can be achieved through combinations of creation and annihilation operators. Expand the interaction density, $ I(x) $, as a sum of terms of the form $
i(x)=|\underline{x}\rangle_1 \ldots |\underline{x}\rangle_m \langle \underline{x}|_{m+1}\ldots\langle \underline{x}|_{n}$ where $ |\underline{x}\rangle_j $ and $ \langle \underline{x}|_j $ are creation and annihilation operators for the particles and antiparticles in the interaction. $i(x)$ can be represented diagrammatically as a vertex or node (figure 1). The diagram is time-ordered from bottom to top so that the lines above the node correspond to creation operators, and those below the node correspond to annihilation operators.
\begin{figure}[h]
	\centering
		\includegraphics[width=0.5\textwidth]{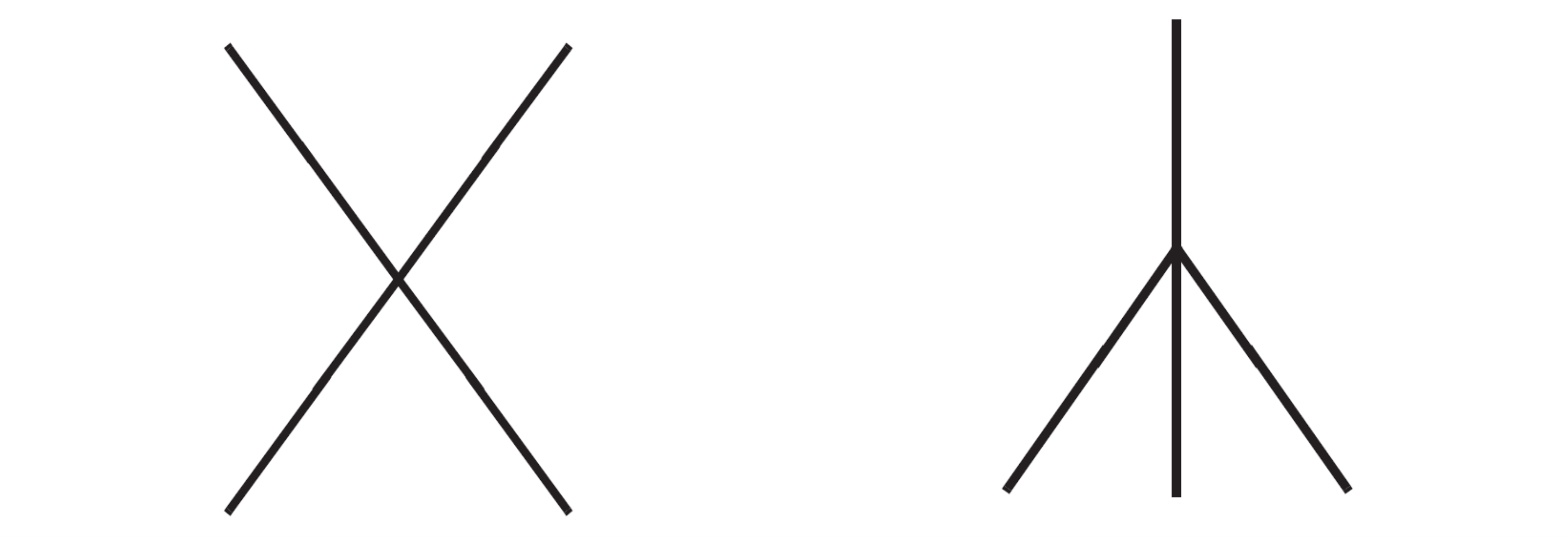}
	\caption{Interaction terms showing a) two creation and two annihilation operators, and b) one creation and three annihilation operators.}
	\label{Fig:1}
\end{figure}

The perturbation expansion for $ \langle g(t)|U(t)|f(0) $ generates a braket between each annihilation operator, $ \langle\underline{x}|_i $, and every earlier creation operator,  $ |\underline{x}\rangle_j $, and every particle in $ |f(0)\rangle $, and a braket between every creation operator,  $ |\underline{x}\rangle_i $, and every particle in the final state, $ \langle g(t)| $. All other brakets are zero. These brakets can be represented graphically by connecting corresponding vertices (figure 2). Lines representing particles are shown with arrows from bottom to top, and lines representing antiparticles with arrows from top to bottom. Then the $ n^{th} $ term of the perturbation expansion is a sum of terms, each represented as a time-ordered graph containing $n$ vertices.
\begin{figure}[h]
	\centering
		\includegraphics[width=0.5\textwidth]{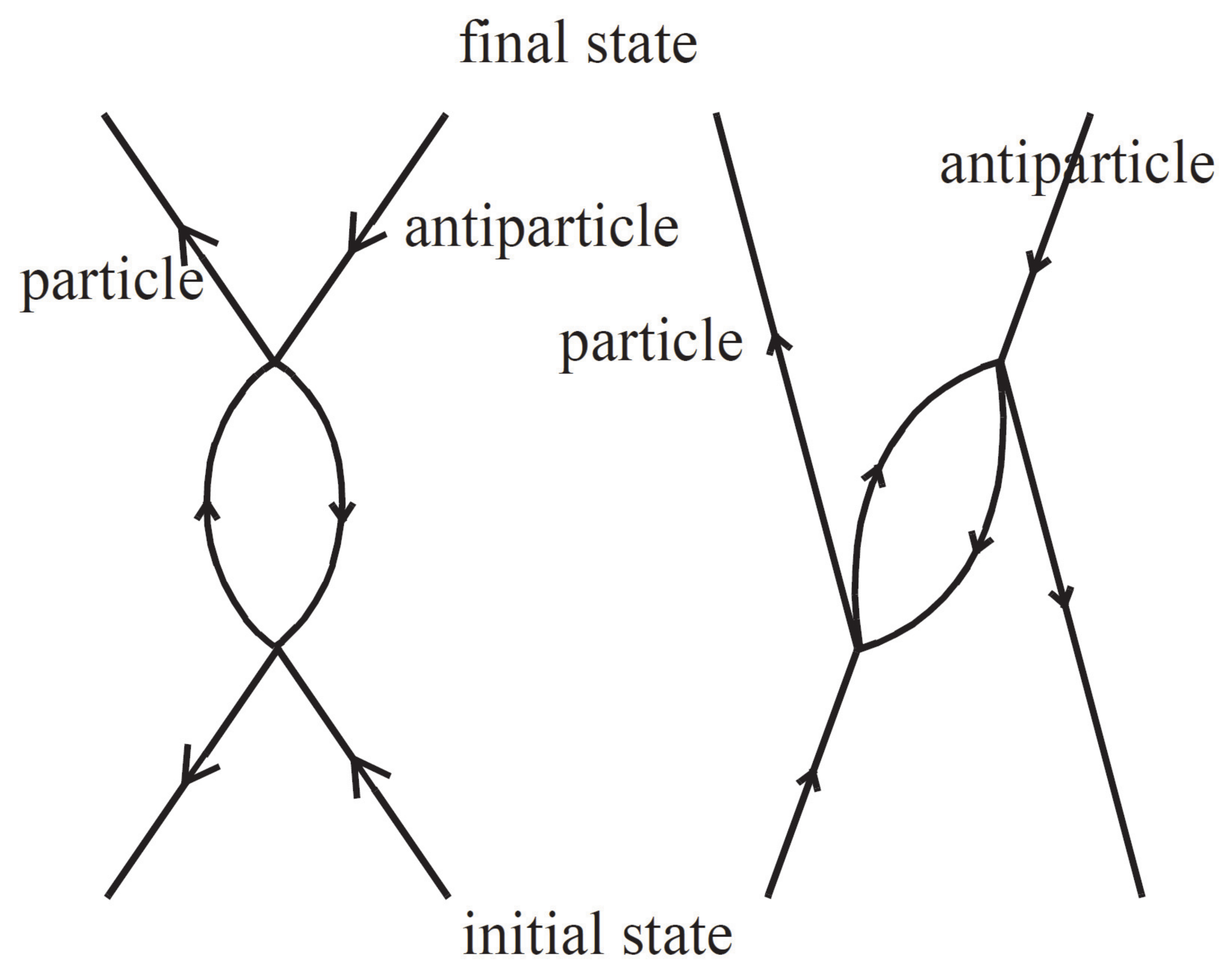}
	\caption{Time-ordered diagrams for two interactions.}
	\label{Fig:2}
\end{figure}
\subsection{The locality condition}\label{sec:4.5}
\begin{flushleft}
\textbf{Definition:} Let $ \lbrace t_{j} \in \mathrm{T}:j=1, \ldots,n \rbrace $  be an unordered set of times in $ \mathrm{T} $. Let $ \pi $ be the permutation such that $ t_{\pi(1)} > t_{\pi(2)} > \ldots > t_{\pi(n)} $. Then the \textbf{time-ordered product}, $\mathcal{T}$, is
\begin{equation}\label{Eq:4.5.1}
\mathcal{T} \lbrace I(t_1)\ldots I(t_1)\rbrace = I(t_{\pi(1)})\ldots I(t_{\pi(n)}).
\end{equation}

\end{flushleft}Hence, we can write the perturbation expansion \eqref{Eq:4.3.6}
\begin{equation}\label{Eq:4.5.2}
U(t_{n})= 1+ \sum\limits_{k=1}^{n} \frac{(-i\chi^4)^k}{k!} \sum\limits_{i_k\ne i_{k-1},i_{k-2},\ldots,i_{1}} \ldots  \sum\limits_{i_2\ne i_1}  \sum\limits_{i_1=0}^{n-1} \mathcal{T}\lbrace I(x_{i_k}) \ldots I(x_{i_2})I(x_{i_1}) \rbrace.
\end{equation}
It will be observed that in the limit $ \chi\rightarrow 0 $ this reduces to the standard integral form (appendix \ref{Ap:G}), and that, because equal time products are excluded, improper integrals must be used. In consequence care is needed in the order of taking limits.\\
\\
\textbf{Theorem:} (\textit{Locality}) For any $x$, $y$, such that $x - y$ is space-like, the commutator of the interaction densities at $ x $ and $ y $ vanishes,  $[I(y), I(x)]$ = 0.\\
\\
\textit{Proof:} Under Lorentz transformation, the order of interactions, $ I(x_i) $, $ I(x_j) $ can be changed in the time-ordered product whenever $ x_i - x_j $ is space-like. Under the condition that the initial and final kets are stable states of free particles, as in scattering experiments, the calculation of probabilities cannot be affected. Locality follows immediately.

\section{Field theory}\label{sec:5}
\subsection{The Dirac field operator}\label{sec:5.1}
The interaction of a particle will be modelled by the annihilation of the old state of the particle and the creation of a new state. Since we can measure the position of an electron, it must be possible to form a projection operator for position at given time,
\begin{equation}\label{Eq:5.1.1}
X_{\mathrm{P}}(x) = |x\rangle \langle x|,
\end{equation}
from the Hamiltonian density and a suitable configuration of matter (more accurately, $ X_{\mathrm{P}}(x) $ is summed over a small range of positions depending on the resolution of measurement). The interpretation of antiparticles as negative energy particles going backwards in time means that the annihilation of a negative energy particle appears as the creation of a (positive energy) antiparticle, so that antiparticle annihilation operators appear symmetrically in the interaction operator in a sum with creation operators. This motivates the definition of the Dirac field operator.\\
\\
\textbf{Definition:} The \textbf{Dirac field operator} annihilates an electron or creates a positron,
\begin{equation}\label{Eq:5.1.2}
\psi_{\alpha}(x) = |\bar{x},\alpha\rangle + \langle \underline{x},\alpha|.
\end{equation}

There are strong reasons, based on locality, for thinking that interaction operators are products of field operators with this form. The Hermitian conjugate of a quantum field operator, has the reverse effect, creating a particle or annihilating an antiparticle. The observable quantity, current density, uses the Dirac adjoint, so we expect the Dirac adjoint operators to appear alongside field operators in the Hamiltonian density.\\
\\
\textbf{Definition:} The \textbf{Dirac adjoint} of the annihilation operator $ \langle \underline{x},\alpha| $ creates a particle, 
\begin{equation}\label{Eq:5.1.3}
|\hat{\underline{x}},\alpha\rangle = \sum\limits_{\mu} |\underline{x},\mu\rangle \gamma_{\mu \alpha}^0. 
\end{equation}
The bound on the momentum space integral does not affect covariance, since the operators act on kets having wave functions with bounded support in momentum space.\\
\\
\textbf{Definition:} The \textbf{Dirac adjoint} of the antiparticle creation operator $ |\bar{x}, \alpha \rangle $ annihilates an antiparticle,
\begin{equation}\label{Eq:5.1.4}
\langle\hat{\bar{x}},\alpha | = \sum\limits_{\mu} \langle\bar{x},\mu| \gamma_{\mu \alpha}^0. 
\end{equation}
\textbf{Definition:} The \textbf{Dirac adjoint} of the field operator creates a particle or annihilates an antiparticle,
\begin{equation}\label{Eq:5.1.5}
\hat{\psi}_{\alpha}(x) = \psi^{\dagger}_{\mu} \gamma_{\mu \alpha}^0 =     |\hat{\underline{x}},\alpha\rangle + \langle \hat{\bar{x}},\alpha|.
\end{equation}

\subsection{Locality of Dirac field operators}\label{sec:5.2}
Since Dirac particles are Fermions we have anticommutation relations for the Dirac field operator. 
\begin{equation}\label{Eq:5.2.1}
\lbrace \psi_{\alpha}(x) , \psi_{\beta}(y) \rbrace = \lbrace \hat{\psi}_{\alpha}(x), \hat{\psi}_{\beta}(y)\rbrace = 0.
\end{equation}
Dirac field operators will appear in pairs in the Hamiltonian density in such a way as to ensure commutation relations required of the locality condition.\\
\\
\textbf{Theorem:} The equal time anticommutation relations for the Dirac field and Dirac adjoint and obey:
\begin{equation}\label{Eq:5.2.2}
\lbrace \psi_{\alpha}(x) , \hat{\psi}_{\beta}(y) \rbrace_{x^0=y^0} = \gamma_{\alpha\beta}^0 \delta(\boldsymbol{x}-\boldsymbol{y}).
\end{equation}
\\
\textbf{Theorem:} (locality) The anticommutation relation for the Dirac field and the Dirac adjoint is zero outside the light cone.\\
\\
\textit{Proof:} appendix \ref{Ap:H}.

\subsection{The current density observable}\label{sec:5.3}
For electrons, current is an observable quantity. Since measurement is always the result of interactions between matter, a Hermitian operator, $ j $, whose expectation is the classical electrical current must appear in the Hamiltonian density. To ensure that locality is satisfied, current is composed of Dirac field operators.\\
\\
\textbf{Postulate:} The \textbf{current density observable} is $ j^{\alpha}(x) = :  \hat{\psi}(x) \gamma^{\alpha} \psi(x) : $, where the colons denote normal ordering (i.e. creation operators are placed to the left of annihilation operators).\\
\\
Then, given the particle ket $ |f\rangle $ in $ \mathbb{H}^{1} $ with $ f(x)= \langle x|f \rangle $,
\begin{equation}\label{Eq:5.3.1}
\langle j^a(x) \rangle = \langle f | j^a(x) |f\rangle = \langle f |: (|\hat{\underline{x}}\rangle + \langle \hat{\bar{x}}| )\gamma^a (|\bar{x}\rangle + \langle \underline{x}|:) |f\rangle = \hat{f}(x)\gamma^a f(x),
\end{equation}
in agreement with current density for a single particle state (section \ref{sec:3.6}). For antiparticle states, spin indices are transposed. Transposition is equivalent to pre- and post-multiplying $ \gamma $ by a matrix, with ones in the trailing diagonal and zeros elsewhere. This has the effect of reversing the order of the spin indices. Thus, a negative energy spin down electron will appear as a positive energy spin up positron.

\subsection{Photons}\label{sec:5.4}
We seek to introduce interactions between particles, such that the interaction operator has an invariant form. Since the current density observable, $ j^{a}(x) $, is a vector, a covariant theory can be found by contracting it with another Hermitian vector operator, $ A^{a}(x) $. The possibilities are severely restricted. The natural and simplest thing to try is to introduce a particle with a spin index which transforms as a vector, and which is its own antiparticle, i.e. its creation and annihilation operators appear in the same field operator. Vector particles may have non-zero mass, but empirical evidence is that this is not so for the photon at the limit of experimental accuracy. Zero mass is assumed.\\
\\
\textbf{Postulate:} The \textbf{photon field operator} is $  A^{a}(x)=|\underline{x},a\rangle + \langle \underline{x}, a| $.\\

The photon field operator creates or annihilates a photon, in accordance with the empirical fact that photons are created or destroyed in interaction. We cannot, therefore, talk of measurements of the position of a photon (a position observable), but only of measurement of the position at which it was annihilated, or the position at which it was created. $x$ is not the position of a photon, but rather the position at which a charged particle would be found to have emitted or absorbed a photon if a measurement were carried out. This requires that we extend the formal rules introduced in section 2.5 and section 2.6.\\
\\
\textbf{RULE VIIIa.} $ |x,a\rangle = |\underline{x}, a\rangle | \rangle $ is the formal conditional clause \textit{``If a measurement had found the creation of a photon at $ x $, \ldots''}.\\
\\
\textbf{RULE VIIIb.}  $ \langle x,a | = \langle|\langle\underline{x}, a | $ is the formal consequent clause \textit{``\ldots, then a measurement would find the annihilation of a photon at $ x $''}.\\
\\
\textbf{RULE VIIIc.} The \textbf{photon position function}, $ \langle x,a|f\rangle = \langle A^a(x)|f\rangle $  is the formal statement, \textit{``if  $ |f\rangle $ were known from previous measurement(s), then, another measurement would find the annihilation of the photon at $ x $''}.\\
\\
Since photons are always created or annihilated in interaction, and cannot be in eigenkets of a position observable, we do not require that $ |x,a\rangle $ are orthogonal. 

\subsection{Plane wave photon kets}\label{sec:5.5}
For momentum $ \boldsymbol{p}$, define a longitudinal unit 3-vector, $  \boldsymbol{w}(\boldsymbol{p}, 3) = \boldsymbol{p} / |\boldsymbol{p}| $ and orthogonal transverse unit 3-vectors $ \boldsymbol{w}(\boldsymbol{p}, 1)$ and 	$ \boldsymbol{w}(\boldsymbol{p}, 2)$ such that, for $ r = 1, 2, 3 $, 	$ \boldsymbol{w}(\boldsymbol{p}, r) \cdot \boldsymbol{w}(\boldsymbol{p}, s) = \delta_{rs}$, and define normalised \textbf{spin vectors}, $ w(\boldsymbol{p}, 0) = (1, \boldsymbol{0})$ and $ w(\boldsymbol{p}, r) = (0, \boldsymbol{w}(\boldsymbol{p}, r))$ for $ r = 1, 2, 3$.\\
\\
\textbf{Definition:} For momentum $ \boldsymbol{p} $, and $ r = 0, 1, 2, 3 $, the \textbf{photon plane wave ket}, $ |\boldsymbol{p},r\rangle $, in $ \mathbb{H}^1 $ is given by the wave function, 
\begin{equation}\label{Eq:5.5.1}
\langle x|\boldsymbol{p},r \rangle = \lambda (|\boldsymbol{p}|,r) w(\boldsymbol{p},r)e^{-ix \cdot p},
\end{equation}
where $ p^{2} = 0$ (the mass shell) and $ \lambda(|\boldsymbol{p}|,r) $ is a scalar, to be determined.\\
\\
The scalar, $ \lambda $, is required because the kets $ |x\rangle = |\underline{x} \rangle |\rangle $  refer to the hypothetical measurement of position of the electron which emits a photon, not to the position at which a photon can be measured. Direction is determined by the distribution of matter, not by fundamental assumption, so $ \lambda $ depends only on the magnitude of $ \boldsymbol{p} $.

Since momentum is a conserved quantity (appendix \ref{Ap:J.4}), a photon created with a given momentum is annihilated with the same momentum and it is possible to talk about the measured momentum of a photon state. We therefore require that plane wave kets are an orthogonal basis,
\begin{equation}\label{Eq:5.5.2}
\langle \boldsymbol{q},s| \boldsymbol{p},r \rangle = \eta(r)\delta_{rs}\delta(\boldsymbol{p}-\boldsymbol{q}),
\end{equation}
where $ \eta(0) = -1 $ and $ \eta(r) = 1 $ for $ r = 1, 2, 3 $. The minus sign from $\eta(0) $ does not alter the expansion of the inner product for an orthonormal basis. The braket for the photon is
\begin{equation}\label{Eq:5.5.3}
\langle g| f \rangle = \sum\limits_{r=0}^{3}\int_{\mathrm{M}}d^3\boldsymbol{p}\, \langle g|\boldsymbol{p}, r \rangle \langle \boldsymbol{p}, r|f\rangle.
\end{equation}
The resolution of unity takes the form,
\begin{equation}\label{Eq:5.5.4}
1 = \sum\limits_{r=0}^{3}\int_{\mathrm{M}}d^3\boldsymbol{p}\, |\boldsymbol{p}, r \rangle \langle \boldsymbol{p}, r|.
\end{equation}
We do not have that $ \langle f|f \rangle > 0 $ for all $ |f\rangle \in \mathbb{H}^1 $; the braket is not positive definite, in apparent conflict with the calculation of probabilities. In practice, we only need to generate probabilities for \textit{observations}, not for the entire space of photon kets. Since probabilities must be positive, we impose the condition that, in observations on the photon, there is no polarisation between time-like and longitudinal states,
\begin{equation}\label{Eq:5.5.5}
\langle \boldsymbol{p},0| f \rangle = \langle \boldsymbol{p},3| f \rangle,
\end{equation}

Physically, any polarisation breaking \eqref{Eq:5.5.5} would need to be caused by an interaction creating that polarisation. Clearly, no such interaction is known or possible. It is seen that if one starts with Coulomb gauge (as used in many standard approaches), then \eqref{Eq:5.5.5} remains true after Lorentz transform, while Coulomb gauge does not. Using \eqref{Eq:5.5.5}, probabilities for the observation time-like and longitudinal states are zero. The braket reduces to
\begin{equation}\label{Eq:5.5.6}
\langle g| f \rangle = \sum\limits_{r=1}^{2}\int_{\mathrm{M}}d^3\boldsymbol{p} \,\langle g|\boldsymbol{p}, r \rangle \langle \boldsymbol{p}, r|f\rangle,
\end{equation}
which is positive semidefinite, as required for the probability interpretation. It will be seen that all four polarisation states are required for the derivation of the Lorentz force (section \ref{sec:6.3}). We can conclude that the unobservable states have a real effect, and represent real particles, but the probability interpretation allows only the direct observation of a subspace containing the two transverse polarisation states, on which the inner product is positive definite. Since the braket is invariant under the addition of a light-like polarisation state, light-like polarisation cannot be determined from experimental results.

We require that the probability for the creation of a photon at $x$ and its annihilation at $y$ is invariant. Observe that
\begin{equation}\label{Eq:5.5.7}
 \sum\limits_{r=0}^{3}\eta(r)w^a(\boldsymbol{p}, r)w^b(\boldsymbol{p}, r) = - g^{ab}.
\end{equation}
Then, setting
\begin{equation}\label{Eq:5.5.8}
\lambda(|\boldsymbol{p}|,r) = (\tfrac{1}{2\pi})^{3/2}\tfrac{1}{\sqrt{2p^0}}
\end{equation}
gives
\begin{equation}\label{Eq:5.5.9}
\begin{split}
\langle \underline{x},a|\underline{y}, b \rangle &= 
\sum\limits_{r=0}^{3}\int_{\mathrm{M}}d^3\boldsymbol{p}\, \langle \underline{x},a|\boldsymbol{p}, r \rangle \langle \boldsymbol{p}, r|\underline{y}, b\rangle \\ &= 
\sum\limits_{r=0}^{3}\int_{\mathrm{M}}d^3\boldsymbol{p}\, \eta(r)\lambda(|\boldsymbol{p}|,r)\lambda(|\boldsymbol{p}|,r)w^a(\boldsymbol{p}, r)w^b(\boldsymbol{p}, r) e^{-ip \cdot(x-y)}\\ &= 
- \frac{g^{ab}}{8\pi^3}\int_{\mathrm{M}}\frac{d^3\boldsymbol{p}}{2p^0}\, e^{-ip \cdot(x-y)}\\ &=
- \frac{g^{ab}}{8\pi^3}\int_{\mathrm{M}}d^4p\, e^{-ip \cdot(x-y)}\delta(p^2),
\end{split}
\end{equation}
which is covariant because quantum covariance applies to the momentum space integral (see section \ref{sec:3.2}).

\subsection{Evolution of photon kets}\label{sec:5.6}
We may expand $|x,a\rangle $ using plane waves,
\begin{equation}\label{Eq:5.6.1}
\begin{split}
|x, a \rangle
&= \sum\limits_{r=0}^{3}\int_{\mathrm{M}}d^3\boldsymbol{p}\, |\boldsymbol{p}, r \rangle \langle \boldsymbol{p}, r|x, a\rangle\\
&= (\tfrac{1}{2\pi})^{3/2} \sum\limits_{r=0}^{3}\int_{\mathrm{M}}\frac{d^3\boldsymbol{p}}{\sqrt{2p^0}}\,w^a(\boldsymbol{p}, r)e^{ix\cdot p} |\boldsymbol{p}, r \rangle 
\end{split}
\end{equation}
Then the wave function for the ket $ |f\rangle $ is
\begin{equation}\label{Eq:5.6.2}
f^a(x) = (\tfrac{1}{2\pi})^{3/2} \sum\limits_{r=0}^{3}\int_{\mathrm{M}}\frac{d^3\boldsymbol{p}}{\sqrt{2p^0}}\,w^a(\boldsymbol{p}, r)e^{-ix\cdot p} \langle \boldsymbol{p}, r|f \rangle
\end{equation}
The Klein-Gordon Equation, $ \partial^2f^a=0 $, is seen here as a vector identity, expressing the mass shell condition for a zero mass particle,  not as an equation of motion. Since the probability of the annihilation of a particle at $ x $ given its creation at $ y $ is the same whenever it is calculated, Stone's theorem \cite{Stone} (appendix \ref{Ap:B}) requires a first order equation, which is found by differentiating the wave function,
\begin{equation}\label{Eq:5.6.3}
\partial_af^a=0.
\end{equation}

\subsection{The photon field operator}\label{sec:5.7}
The creation operator for a plane wave ket is given by $ |\underline{\boldsymbol{p}},r\rangle |\rangle = |\boldsymbol{p},r\rangle $. Expanding the photon field operator, section \ref{sec:5.4},
\begin{equation}\label{Eq:5.7.1}
A^a(x) = \sum\limits_{r=0}^{3}\eta(r)\int_{\mathrm{M}}\frac{d^3\boldsymbol{p}}{\sqrt{2p^0}}\, (e^{ix\cdot p} |\underline{ \boldsymbol{p}}, r \rangle +  e^{-ix\cdot p}\langle \underline{ \boldsymbol{p}}, r| )  w^a(\boldsymbol{p}, r).
\end{equation}
\textbf{Theorem:} The photon field operator satisfies $ \partial^2 A^a =0 $.\\
\\
\textit{Proof:} Differentiate $ A^{a} $ twice.\\
\\
\textbf{Theorem:} For physical states, the Gupta-Bleuler gauge condition is satisfied: 
\begin{equation}\label{Eq:5.7.2}
\partial_aA^a|f\rangle=0.
\end{equation}
\textit{Proof:} Differentiate and use \eqref{Eq:5.5.5}.

\subsection{The locality condition for photons}\label{sec:5.8}
Photons are Bosons, obeying commutation relations,
\begin{equation}\label{Eq:5.8.1}
\begin{split}
[A^a(x),A^b(y)] &= [ |\underline{x},a\rangle+\langle \underline{x},a| , |\underline{y},b\rangle+\langle \underline{y},b| ]\\
&= \langle \underline{x},a|\underline{y},b\rangle - \langle \underline{y},b|\underline{x},a\rangle\\
&= -\frac{g^{ab}}{8\pi^3} \int_{\mathrm{M}}\frac{d^3\boldsymbol{p}}{2p^0}\, (e^{-i(x-y)\cdot p} - e^{i(x-y)\cdot p}).
\end{split}
\end{equation}
Substituting $ \boldsymbol{p}\rightarrow -\boldsymbol{p} $ in the second term gives the equal time commutator,
\begin{equation}\label{Eq:5.8.2}
[A(x),A(y)]_{x^0=y^0}=0.
\end{equation}
The commutator \eqref{Eq:5.8.1} is covariant because \eqref{Eq:5.5.9} is covariant. Hence it is zero outside the light cone. Because the photon commutator vanishes, the time evolution of the expectation of the photon field is trivial. Physical laws depend on derivatives of the photon field.\\
\\
\textbf{Theorem:} The commutator for the photon field and its derivative is zero outside the light cone (locality), and the equal time commutator obeys:
\begin{equation}\label{Eq:5.8.3}
[\partial_iA^a(x),A^b(y)]_{x^0=y^0} = - i \delta_i^0 g^{ab}\delta(\boldsymbol{x}-\boldsymbol{y}).
\end{equation}
\textit{Proof:} Differentiating,
\begin{equation}\label{Eq:5.8.4}
\partial \langle \underline{x},a|\underline{y},b\rangle = -\frac{g^{ab}}{8\pi^3} \int_{\mathrm{M}}\frac{d^3\boldsymbol{p}}{2p^0}\,ip e^{-i(x-y)\cdot p},
\end{equation}
and
\begin{equation}\label{Eq:5.8.5}
\partial \langle \underline{y},b|\underline{x},a\rangle = \frac{g^{ab}}{8\pi^3} \int_{\mathrm{M}}\frac{d^3\boldsymbol{p}}{2p^0}\, ip e^{i(x-y)\cdot p}.
\end{equation}
Substitute $ \boldsymbol{p} \rightarrow -\boldsymbol{p} $ at $ x^0=y^0 $. Then, for $ i=1,2,3 $,
\begin{equation}\label{Eq:5.8.6}
[\partial_iA^a(x),A^b(y)]_{x^0=y^0} = \partial_i \langle \underline{x},a|\underline{y},b\rangle - \partial_i \langle \underline{y},b|\underline{x},a\rangle =0,
\end{equation}
and, for the time-like component,
\begin{equation}\label{Eq:5.8.7}
[\partial_0A^a(x),A^b(y)]_{x^0=y^0} =  -i\frac{g^{ab}}{8\pi^3} \int_{\mathrm{M}}d^3\boldsymbol{p}\, e^{i(\boldsymbol{x}-\boldsymbol{y})\cdot \boldsymbol{p}} = - i  g^{ab}\delta(\boldsymbol{x}-\boldsymbol{y}).
\end{equation}
\eqref{Eq:5.8.4} and \eqref{Eq:5.8.5} are covariant because \eqref{Eq:5.5.9} is covariant. So, $ [\partial_0A^a(x),A^b(y)] $ is covariant and vanishes outside the light cone because it vanishes for$ x^0=y^0 $ .

\section{Electromagnetism}\label{sec:6}
\subsection{The interaction density for qed}\label{sec:6.1}
The photon field operator, $ A(x) $, is Hermitian. It is natural to ask whether its expectation is a classical quantity. Classical quantities are determined from measurement, i.e. through interaction with other matter. They must therefore be describable, at least in principle, in terms of the interaction density and the configuration of matter. If  $\langle A(x)\rangle $ is classical, $ A(x) $ must appear in the interaction density (since it cannot be formed as a composition of simpler operators obeying locality). Qed uses the intuitively appealing minimal interaction, in which single photons are emitted and absorbed by electrons.\\
\\
\textbf{Postulate:} The \textbf{Hamiltonian density for qed} is 
\begin{equation}\label{Eq:6.1.1}
I(x)=ej^a(x)A_a(x) 
\end{equation}
where $ e $ is an experimentally determined constant, \textbf{charge}, and $ j $ is the current density observable, $ j^{\alpha}(x) = :  \hat{\psi}(x) \gamma^{\alpha} \psi(x) : $ (section \ref{sec:5.3}).\\
\\
To establish that $\langle A(x) \rangle$ is the classical electromagnetic field, it is necessary to establish the Lorentz force law (section \ref{sec:6.3}) and Maxwell's equations (section \ref{sec:6.4}).\\
\\
\textbf{Theorem:} $\langle A(x) \rangle$ satisfies the Lorenz gauge condition, $\partial_a\langle A^a(x) \rangle =0$.\\
\\
\textit{Proof:} Apply Ehrenfest's theorem (appendix \ref{Ap:D}). By locality the equal time commutator is zero. Using the Gupta-Bleuler gauge condition \eqref{Eq:5.7.2},
\begin{equation}\label{Eq:6.1.2}
\partial_a\langle A^a(x) \rangle = \langle \partial_a A^a(x) \rangle = 0
\end{equation}
The Lorenz gauge condition fixes gauge up to the unobservable light-like polarisation. In classical electrodynamics one may choose a different gauge without affecting predictions, but here Lorenz gauge is fixed by the Gupta-Bleuler gauge condition, which in turn arises from the absence of polarisation between time-like and longitudinal states \eqref{Eq:5.5.5}, required to preserve the probability interpretation.

\subsection{Momentum in the interacting theory}\label{sec:6.2}
In the absence of interactions, there is no issue with local gauge freedom. The phase of an electron wave function is fixed at the point of creation and becomes simply the global symmetry of the one particle theory, in which kets can be multiplied by constant phase without altering their meaning in formal language. When interactions are introduced the result is that the evolution of the wave function does not match the evolution of the field operator which created it, and which is defined on the non-interacting space. A difficulty arises because the momentum observable in the non-interacting theory,
\begin{equation}\label{Eq:6.2.1}
P^a=i \sum\limits_{D}|x\rangle\partial^a\langle x| = i\partial^a
\end{equation}
extracts the frequency and wavelength of the wave function. We would like to use Ehrenfest's theorem (appendix \ref{Ap:D}) to calculate the classical force due to the interaction, by differentiating the expectation of momentum,
\begin{equation}\label{Eq:6.2.2}
\frac{d}{dt}\langle P^a \rangle = \langle \frac{d}{dt} P^a \rangle + i \langle [H,P^a]\rangle,
\end{equation}
but kets evolve according to the full Hamiltonian, whereas the creation operators are defined on the Fock space of non-interacting particles, and create kets obeying the Dirac equation. There is a real phase shift corresponding to change in momentum, which must be distinguished from the arbitrary phase in the definition of field operators.

To ensure that creation operators and states evolve identically, we define the \textbf{field picture}, using a simplified form of the Foldy-Wouthuysen transformation \cite{Foldy} 
which ignores spin,
\begin{equation}\label{Eq:6.2.3}
|f_{\mathrm{F}(t)}\rangle = e^{-iH_{\mathrm{I}}t} | f(t)\rangle =  e^{iH_{\mathrm{0}}t}| f(0)\rangle.
\end{equation}
In the field picture kets evolve as in the Schr\"{o}dinger picture for non-interacting particles. The momentum operator in the field picture is 
\begin{equation}\label{Eq:6.2.4}
P_{\mathrm{F}}^a = e^{-iH_{\mathrm{I}}t} i\partial^a e^{iH_{\mathrm{I}}t}.
\end{equation}
In the semi-classical correspondence, for small $t$, evolution may be treated as a perturbation to the evolution of a non-interacting particle, in which the interaction Hamiltonian is replaced with its expectation. For a classical particle with position $x$ and velocity $\dot{x} $, the classical current is $ J=-e\dot{x} $. The expectation of the interaction Hamiltonian is
\begin{equation}\label{Eq:6.2.5}
\langle H_{\mathrm{I}} \rangle = J \cdot \langle A(x) \rangle = -e\dot{x}\cdot \langle A(x) \rangle .
\end{equation}
Replacing the interaction Hamiltonian with its expectation, the momentum operator in the field picture is
\begin{equation}\label{Eq:6.2.6}
P_{\mathrm{F}}^a = e^{ie\dot{x}\cdot \langle A(x) \rangle t} i\partial^a e^{-ie\dot{x}\cdot \langle A(x) \rangle t} = i\partial^a - e \langle A^a(x) \rangle .
\end{equation}
Thus the expectation, $ \langle A(x) \rangle $, of the operator which creates and annihilates photons, acts in the manner of a classical vector field, modifying energy and momentum. This is the standard formula for generalised momentum in the presence of a field, but normally it is assumed on phenomenological grounds, whereas here it is been found from theoretical considerations. With the replacement of the momentum operator for non-interacting particles with the corresponding operator taking interactions into account, $ i\partial^a \rightarrow P_{\mathrm{F}}^a =  i\partial^a - e \langle A^a(x) \rangle  $, the Dirac equation, $ (i\gamma^a\partial_a-m)f(x)=0 $, becomes the interacting Dirac equation,
\begin{equation}\label{Eq:6.2.7}
(\gamma^a(i\partial_a - e \langle A_a(x) \rangle) -m)f(x)= 0.
\end{equation}

\subsection{The Lorentz force law}\label{sec:6.3}
Working in the field picture, we have, from Ehrenfest's theorem (appendix \ref{Ap:D}),
\begin{equation}\label{Eq:6.3.1}
\frac{d}{dt}\langle P_{\mathrm{F}}^a \rangle = \langle \frac{d}{dt} P_{\mathrm{F}}^a \rangle + i \langle [H,P_{\mathrm{F}}^a]\rangle.
\end{equation}
In the classical correspondence we use the expectation in place of the interaction:
\begin{equation}\label{Eq:6.3.2}
H = H_0 + H_{\mathrm{I}} \approx  H_0 + \langle H_{\mathrm{I}} \rangle =  H_0 - e\dot{x}\cdot\langle A(x) \rangle.
\end{equation}
Substituting for $ H $ in \eqref{Eq:6.3.1}, using \eqref{Eq:6.2.6}, and dropping the suffix $ \mathrm{F} $ (since expectations are the same in any picture),
\begin{equation}\label{Eq:6.3.3}
\begin{split}
\frac{d}{dt}\langle P^a \rangle &= e\frac{d}{dt}\langle A^a(x) \rangle + i\langle [ H_0 - e \dot{x} \cdot \langle A(x) \rangle, i\partial^a - e \langle A^a(x) \rangle  ] \rangle\\
&= e \frac{d}{dt}\langle A^a(x) \rangle - e\partial^a(\dot{x} \cdot \langle A(x) \rangle),
\end{split}
\end{equation}
where the product rule of differentiation has been used to find the second term. Classical force is defined as the rate of change of momentum, according to Newton's second law,
\begin{equation}\label{Eq:6.3.4}
(\mathrm{Force})^a \equiv \frac{d}{d\tau} (\mathrm{momentum})^a,
\end{equation}
where, by general covariance, $\tau$ is proper time for the matter on which the force acts, not coordinate time defined by a particular observer. The electromagnetic force on a charged particle is evaluated in the rest frame of the particle, in which $ \dot{x} =(1,0,0,0)$ and current is $ J=-e(1,0,0,0) $. Then \eqref{Eq:6.3.3} is
\begin{equation}\label{Eq:6.3.5}
\partial^0 \langle P^a \rangle = e \partial^0 \langle A^a(x) \rangle - e \partial^a \langle A^0(x) \rangle.
\end{equation}
So,
\begin{equation}\label{Eq:6.3.6}
(\mathrm{Force})^a \equiv \frac{d}{d\tau} \langle P^a \rangle = J_0(\partial^a \langle A^0(x)\rangle - \partial^0 \langle A^a(x)\rangle ).
\end{equation}
After Lorentz transformation, this is
\begin{equation}\label{Eq:6.3.7}
(\mathrm{Force})^a \equiv \frac{d}{d\tau} \langle P^a \rangle = J_b(\partial^a \langle A^b(x)\rangle - \partial^b \langle A^a(x)\rangle )=J_bF^{ab},
\end{equation}
where $ F^{ab} $ is the \textbf{Faraday tensor}
\begin{equation}\label{Eq:6.3.8}
F^{ab} = \partial^a \langle A^b(x)\rangle - \partial^b \langle A^a(x)\rangle.
\end{equation}
This establishes the Lorentz force law.

\subsection{Maxwell's equations}\label{sec:6.4}
\begin{flushleft}
\textbf{Theorem:} $ \langle A(x) \rangle $ satisfies Maxwell's equations in Lorenz gauge:
\begin{equation}\label{Eq:6.4.1}
\partial^2 \langle A(x) \rangle = -e \langle j(x) \rangle.
\end{equation}

\end{flushleft}\textit{Proof:} Differentiating the expectation of the photon field twice, using Ehrenfest's theorem (appendix \ref{Ap:D}),
\begin{multline}\label{Eq:6.4.2}
\partial^2 \langle A(x) \rangle = \partial_a \langle \partial^a A(x) \rangle \\= i\langle[H(x),\partial_0 A(x)]\rangle + \langle\partial^2 A(x) \rangle = i\langle [H(x),\partial_0 A(x)]\rangle.
\end{multline}
Using the Hamiltonian density, \eqref{Eq:6.1.1}, $ I(x) = ej(x) \cdot A(x) $,
\begin{equation}\label{Eq:6.4.3}
\partial^2 \langle A(x) \rangle = ie\chi^3 \sum\limits_{y\in\mathrm{D}}\langle [j(y)\cdot A(y),\partial_0 A(x)]\rangle.
\end{equation}
Maxwell's equations in Lorenz gauge \eqref{Eq:6.4.1} follow immediately by applying the equal time commutator for photons \eqref{Eq:5.8.3}.

\section{Finite quantum electrodynamics}\label{sec:7}
\subsection{The Feynman propagator}\label{sec:7.1}
\begin{flushleft}
\textbf{Definition:} Let $ \phi $ be a field operator, $ \phi(x)=|\bar{x}\rangle+\langle\underline{x}| $. The \textbf{Feynman propagator}, or \textbf{contraction} of $ \phi^{\dagger}(y) $ and $ \phi(x) $ is
\begin{equation}\label{Eq:7.1.1}
D(x-y) = \Theta(x^0-y^0)\langle\underline{x}|\underline{y}\rangle \pm \Theta(y^0-x^0)\langle\bar{x}|\bar{y}\rangle,
\end{equation}

\end{flushleft}where $ + $ is used for Bosons and $ - $ for Fermions, and $ \Theta $ is the step function, $ \Theta(t) = 0 $ if $ t\le 0 $, $ \Theta(t) = 1 $ if $t > 0$. \\
\\
Note that $ D(x-y)=0 $ if $x^0=y^0 $. This may be compared with causal perturbation theory \cite{Scharf}
, using the method of Epstein and Glaser  \cite{Epstein}
, in which the step functions are replaced with a $ C^\infty $ switching function which vanishes at $ t=0 $. It is essential that the equal time propagator is made to vanish. The difference is that here we use a discrete sum whereas causal perturbation theory uses a continuous switching function, and while Scharf \cite{Scharf} says (p. 163) \textit{``the switching on and off the interaction is unphysical''}, here the equal time propagator is specifically excluded from the perturbation expansion \eqref{Eq:4.5.2} and can be regarded as a physical constraint meaning that only one interaction takes place for each particle in any instant. The analysis of the origin of ultraviolet divergences is effectively the same as that given in causal perturbation theory and lattice regularization in that the limit is taken after removing the equal point multiplication, but here there is a physical justification in terms of discrete particle interactions. Further discussion of the origin of the ultraviolet divergence is given in appendix \ref{Ap:L}.

The photon propagator can be evaluated as shown in appendix \ref{Ap:K.2}; for $ x^0 \ne y^0 $,
\begin{equation}\label{Eq:7.1.2}
\begin{split}
D_F(x-y) &= \Theta(x^0-y^0)\langle\underline{x}|\underline{y}\rangle \pm \Theta(y^0-x^0)\langle\underline{y}|\underline{x}\rangle^{\mathsf{T}}\\
&=\frac{-ig}{16\pi^4}\lim\limits_{\epsilon\rightarrow 0^+} \int d^4\tilde{p}\,\frac{e^{-i\tilde{p} \cdot (x-y)}}{\tilde{p}^2 + 2ip^0 \epsilon + \epsilon^2},
\end{split}
\end{equation}
where $ g $ is the metric tensor, $ \tilde{p}^0 $ is a dummy variable, $ \tilde{p}= ( \tilde{p}^0,\boldsymbol{p}) $ is a non-vector, and $ \tilde{\boldsymbol{p}} $ is 3-momentum from $ y $ to $ x $; $ \tilde{\boldsymbol{p}}=\boldsymbol{p} $ if $ x^0>y^0 $ and $ \tilde{\boldsymbol{p}}=-\boldsymbol{p} $ if $ y^0>x^0 $. Similarly the propagator for a Dirac particle can be evaluated as shown in appendix \ref{Ap:K.3}; for $ x^0 \ne y^0 $,
\begin{equation}\label{Eq:7.1.3}
\begin{split}
S_F(x-y) &= \Theta(x^0-y^0)\langle\underline{x}|\hat{\underline{y}}\rangle \pm \Theta(y^0-x^0)\langle\hat{\bar{y}}|\bar{x}\rangle^{\mathsf{T}}\\
&=\frac{i}{16\pi^4}\lim\limits_{\epsilon\rightarrow 0^+} \int d^4\tilde{p}\,\frac{(\tilde{p}\cdot \gamma + m )e^{-i\tilde{p} \cdot (x-y)}}{\tilde{p}^2 - m + i \epsilon}.
\end{split}
\end{equation}

We may now derive Feynman rules following Dyson's calculation (appendix \ref{Ap:J}), but we observe that the integral form of the perturbation expansion \eqref{Eq:G.5} contains improper integrals, and that the limit should not be taken until \textit{after} calculation of each diagram, and that an energy cut-off is automatically introduced by a finite lattice. The most straightforward way to determine the effect of setting $ D(x-y)=0 $ at $ x^0=y^0 $ is to consider the non-perturbative solution. This allows us to impose regularisation conditions on the propagator at low energies, that it is independent of lattice spacing $ \chi $ to first order, and that the renormalised mass and charge adopt their bare values, since the derivations of the Lorentz force law (section \ref{sec:6.3}) and Maxwell's equations (section \ref{sec:6.4}) show that the bare values are physical values.

\subsection{The Landau pole}\label{sec:7.2}
Given the use of improper integrals and nonuniform convergence in the perturbation expansion, it is essential to convergence that all limits are carried out and that the correct order of taking limits is observed. This is precisely what is done in standard regularisation procedures, such as the method of Epstein and Glaser \cite{Epstein} 
and lattice regularisation. The difference between the model described here and lattice regularisation is that here the lattice is a property of measurement, which is necessarily discrete and affects the description of physics but has no impact on underlying physics. The lattice not a property of space, and so it is not required to treat it in the limit of small lattice spacing.

Regularisation ensures that diagrams are finite, but does not ensure the convergence of the perturbation expansion itself. The divergence of the perturbation expansion is usually understood from the Landau pole. Neither causal perturbation theory nor lattice regularization remove the Landau pole, although, since by energy conservation it can only appear on an internal line under a loop integral, it is not clear that it generates a divergence after integration (given a correct order of taking limits). The Landau pole is seen in e.g. eq. (7.96) of Peskin \& Schroeder \cite{Peskin}
. This equation is derived after writing the photon two point function as the sum of one particle irreducible representations (arguments to eq. (7.74) \cite{Peskin}),
\begin{figure}[h]
	\centering
		\includegraphics[width=0.8\textwidth]{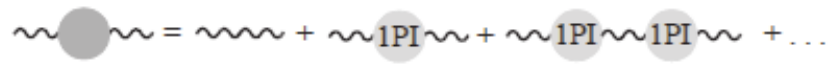}
\end{figure}\\
and treating this as the sum of a geometric progression. But this is a reordering of a series for which we do not have unconditional convergence, and the Riemann series theorem states that for a conditionally convergent series the terms can be reordered such that the series converges to any given value, or even diverges. It is thus not clear that the Landau pole is real.

In this discrete model the perturbation expansion, \eqref{Eq:4.5.2}, terminates after a finite number of terms and cannot generate a divergence. Since the Landau pole does not appear at any finite lattice spacing, and does not appear in the non-perturbative solution, it is not a divergence of the theory. Indeed, if a minimum discrete unit of time, $ \chi $, is a fundamental property of nature, the perturbation expansion necessarily terminates after a finite number of terms, and divergences cannot appear. 

\subsection{The Dyson Instability}\label{sec:7.3}
Dyson \cite{Dyson} 
presented an argument showing \textit{``\ldots that all the power-series expansions currently in use in quantum electrodynamics are divergent after the renormalization of mass and charge''}. He observed that in a fictitious world with a negative fine structure constant, $ \alpha = \frac{e^2}{4\pi} < 0 $, like charges attract, and opposite charges repel. Then a sufficiently large conglomeration of electrons (or positrons) would have lower total energy than the vacuum. He concluded that there must be a finite amplitude for the vacuum to decay into two separate regions, one containing electrons and the other containing positrons. Once such a situation has arisen, the potential due to the charged regions encourages further pair creation, and the runaway decay of the vacuum takes place.

Given infinite time, the runaway decay of the vacuum is a certainty. Thus, when $ \alpha < 0 $ we have zero amplitude, $ f(\alpha) = 0 $, for a typical process studied in perturbation theory such as two particle scattering. We may conclude from analycity that $ f(\alpha) = 0 $ within the radius of convergence. This contradicts the calculation of scattering amplitudes when $ \alpha < 0 $. Dyson concluded that the radius of convergence for any perturbation series in convergence is zero. The decay of the vacuum involves many particles, and Dyson points out that \textit{``The divergence in no way restricts the accuracy of practical calculations that can be made with the theory, but raises important questions of principle concerning the nature of the physical concepts upon which the theory is built''}.

The present treatment allows that in real experiments time is not infinite, and that interactions are discrete. Then the runaway decay of the vacuum is not a certainty in the fictitious model, we do not have  $ f(\alpha) = 0 $ for $ \alpha < 0 $, and non-zero analytic continuation is possible for $ \alpha > 0 $. In fact, since the expansion, \eqref{Eq:4.5.2}, is finite, there are no questions of principle regarding convergence, and we merely observe, with Dyson, that for practical purposes the calculation of lower order terms is unaffected. Physical quantities derived from \eqref{Eq:4.5.2} are well-defined and calculable, but the standard perturbation expansion uses infinite time, both in its definition as a series, and termwise in the calculation of Feynman rules when a delta function is used to conserve energy at a vertex (\ref{Ap:J.4}). In consequence, the standard (infinite) perturbation expansion is asymptotic to \eqref{Eq:4.5.2}, not equal to it.

\subsection{Interpretation of Feynman diagrams}\label{sec:7.4}
In standard treatments of qed, Feynman diagrams are regarded merely as aids to calculation, not descriptions of underlying structure. By contrast, in this treatment the perturbation expansion is interpreted directly as a quantum-logical statement, meaning that any number of interactions might be found taking place at any time and any position if we were to do a measurement. The sums in the expansion simply represent \textsc{or} between possibilities. $ H_{\mathrm{I}}(x) $ describes the possibility that an interaction might be anywhere, not a quantized ``matter field'' which is, in some sense, everywhere. Similarly, Feynman's path integral, or ``sum over all paths'' has as natural interpretation as a logical \textsc{or} between the possible paths that might be detected if an experiment could be done to trace the path (\textit{not} that a particle passes through all paths in spacetime; e.g. Feynman \cite{Feynman4}
).

The perturbation expansion \eqref{Eq:4.5.2} is a sum of terms representing different numbers of interactions. Sum stands for disjunction. So, the meaning of the perturbation expansion is that we cannot say how many interactions take place in any given physical process. Feynman diagrams give a pictorial representation of the same statement; in a particle interpretation, Feynman diagrams also give a pictorial representation of the fundamental structure of matter. We cannot say what the precise configuration of particle interactions in any given instance, but we represent each possible configuration as a graph and sum over the possibilities, using the interpretation of sum as logical disjunction. Only the topology of lines and vertices is relevant. The paper on which the diagram is drawn has no meaning. Spacetime structure does not appear in Feynman diagrams, except in so far as energy-momentum is four dimensional. Thus Feynman diagrams describe the fundamental structure of a particulate relational model in which only particles exist and in which other properties, including spacetime geometry, emerge from interactions between particles.

\section{Conclusion}\label{sec:8}
Classical electromagnetism and quantum electrodynamics have been shown in a particulate model of relativistic quantum mechanics. This was not found using the quantization of classical quantities or the second quantization of classical fields, but by making formal statements about hypothetical measurement results. In this interpretation, qed is fundamentally a theory of particles, not a theory of fields. Classical quantities are understood as expectations, describing the large scale behaviour of systems of many particles. Wave functions are statements in the subjunctive move describing the possible positions where a particle might be found if an experiment were done.

The inner product is invariant because it is defined on an invariant lattice, determined from the measurement apparatus used to define the reference frame. Quantum covariance does not require the limit of small lattice spacing or large lattice size, although it is required that terms dependent on lattice spacing are negligible in the predictions of the theory.

The photon is seen as a particle, and is responsible for the transmission of the classical electromagnetic force, but since a photon cannot be detected without being annihilated, there is no position operator for a photon. The photon wave function describes possibilities for where a photon may be found to have been annihilated, not for where it is. Likewise, field operators describe possibilities rather than actualities and are the mathematical building blocks for the description of interactions between fundamental particles.

Using finite dimensional Hilbert space, fields are operators, not distributions, and there is no problem of principle in taking products of fields. However, in a discrete model the equal point product, $ \phi^†(x)\phi(x) $, does not appear in the perturbation expansion (section \ref{sec:4.5}). The origin of the ultraviolet divergence in the integral form of the perturbation expansion \eqref{Eq:G.5} is the incorrect order of taking limits for diagrams containing improper integrals. In the correct treatment of improper integrals, a cut-off must be used and the limit must be taken \textit{after} calculation of each diagram. Provided that limits are not taken prematurely, terms containing the equal point multiplication do not appear in the perturbation expansion. The exclusion of these terms removes cut-off dependencies to first order and regularizes the perturbation expansion. The regularization condition is that charge and mass adopt their physical values, as in standard treatments. However, it is seen from the derivations of the Lorentz force law (section \ref{sec:6.3}) and Maxwell's equations (section \ref{sec:6.4}) that bare mass and charge are the physical values.\\
\\

\appendix 
\part*{Appendices}
The arguments in the appendices are known. They included for completeness.\\
\section{Unitarity of U} \label{Ap:A}
Since kets can be chosen to be normalised we may require that $U$ conserves the norm, i.e., for all $|g\rangle \in \mathbb{H}$, $\langle g | U^{\dagger}U |g \rangle = \langle g | g \rangle$. Applying this to $|g\rangle + |f\rangle$,
\begin{equation}\label{Eq:A.1}
		(\langle g | + \langle f |)U^{\dagger}U (|g \rangle + |f \rangle) =(\langle g | + \langle f |) (|g \rangle + |f \rangle).
\end{equation}
By linearity of $U$,
\begin{equation}\label{Eq:A.2}
		(\langle g |U^{\dagger} + \langle f |U^{\dagger}) (U|g \rangle + U|f \rangle) =(\langle g | + \langle f |) (|g \rangle + |f \rangle).
\end{equation}
By linearity of the inner product,
\begin{multline*}
\langle g|U^{\dagger}U |g \rangle + \langle g|U^{\dagger}U |f \rangle + \langle f|U^{\dagger}U |g \rangle + \langle f|U^{\dagger}U |f \rangle \\ = \langle g |g \rangle + \langle g|f \rangle + \langle f |g \rangle + \langle f|f \rangle.
\end{multline*}
\begin{equation}\label{Eq:A.4}
\langle g|U^{\dagger}U |f \rangle + \langle f|U^{\dagger}U |g \rangle  =  \langle g|f \rangle + \langle f |g \rangle .
\end{equation}
Similarly, conservation of the norm of $|g\rangle + i |f\rangle$ gives
\begin{equation}\label{Eq:A.5}
\langle g|U^{\dagger}U |f \rangle - \langle f|U^{\dagger}U |g \rangle  =  \langle g|f \rangle - \langle f |g \rangle .
\end{equation}
Combining \eqref{Eq:A.4} and\eqref{Eq:A.5} shows that $U$ is unitary, i.e. for all $|f\rangle,|g\rangle \in \mathbb{H}$,
\begin{equation}\label{Eq:A.6}
\langle g| U^{\dagger}U |f \rangle =\langle g |f \rangle.
\end{equation}

\section{Stone's theorem}\label{Ap:B}
\textbf{Theorem:} (Marshall Stone, 1932). Let $\left\{ U(t)|t \in \mathbb{R} \right\}$ be a set of unitary operators on a Hilbert space, $\mathbb{H}$, $U(t): \mathbb{H} \rightarrow \mathbb{H}$, such that $U(t+s) = U(t)U(s)$ and 
\begin{equation*}
\forall t_0 \in \mathbb{R}, |f\rangle \in \mathbb{H}, \lim_{t \rightarrow t_0} U_t |f\rangle = U_{t_0} |f\rangle
\end{equation*}
then there exists a unique self-adjoint operator $H$ such that $U(t) = e^{-iHt}$.\\
\\
\textit{Proof:} The derivative of $U$ is
\begin{equation}\label{Eq:B.1}
\begin{split}
\dot{U}(t) &= \lim_{dt\rightarrow 0}\frac{U(t+dt)-U(t)}{dt}= \lim_{dt\rightarrow 0}\frac{U(dt)U(t)-U(t)}{dt} \\ &= \left(\lim_{dt\rightarrow 0}\frac{U(dt)-1}{dt}\right)U(t)= U(t)\left(\lim_{dt\rightarrow 0}\frac{U(dt)-1}{dt}\right).
\end{split}
\end{equation}
This prompts the definition of the Hamiltonian operator:\\
\\
\textbf{Definition:} The \textbf{Hamiltonian} $H: \mathbb{H} \rightarrow \mathbb{H}$ is given by
\begin{equation}\label{Eq:B.2}
H=i\left(\lim_{dt\rightarrow 0}\frac{U(dt)-1)}{dt}\right).
\end{equation}
The Hamiltonian has no dependency on $t$. We have
\begin{equation}\label{Eq:B.3}
\dot{U}(t)=-iHU(t) = -iU(t)H.
\end{equation}
So $-iH= U^{\dagger}\dot{U} = \dot{U}U^{\dagger}$. Since $U$ is unitary, for a small time $dt$,
\begin{equation}\label{Eq:B.4}
1 = U^{\dagger}(t+dt)U(t+dt) \approx [U^{\dagger}(t) + \dot{U}^{\dagger}(t)dt][U(t) + \dot{U}(t)dt].
\end{equation}
Ignoring terms in squares of $dt$, and using $-iH= U^{\dagger}\dot{U}$, $iH^{\dagger}= \dot{U}^{\dagger}U$
\begin{equation}\label{Eq:B.5}
U^{\dagger}(t)U(t) - iH^{\dagger}dt + iHdt \approx 1.
\end{equation}
Using unitarity of $U$, we find that $H$ is Hermitian, $H=H^{\dagger}$. \eqref{Eq:B.3} has solution 
\begin{equation}\label{Eq:B.6}
U(t) = e^{-iHt}.
\end{equation}
\\
\textbf{Corollary:} The wave function satisfies the Schr\"{o}dinger equation
\begin{equation}\label{Eq:B.7}
\partial_0 f(t,x) = -iH f(t,x) .
\end{equation}
\textit{Proof:} Differentiate the wave function using \eqref{Eq:B.3},
\begin{equation}\label{Eq:B.8}
\partial_0 f(t,x)= \langle x | \dot{U} |f(0) \rangle = \langle x | - iHU(t) |f(0) \rangle = \langle x | -iH |f(t) \rangle.
\end{equation}
\\
\textbf{Corollary:} Newton's first law.\\
\\
\textit{Proof:} After replacing 3-vectors with 4-vectors in \eqref{Eq:2.7.2} and imposing the mass shell condition, $E^2 = (p^0)^2 = m^2 + \boldsymbol{p}^2$ for some constant $m$, we find that a plane wave is a solution of the Schr\"{o}dinger equation with  $H=E= \text{const}$. Thus momentum, $p$, does not change in time for a non-interacting particle.

\section{The Derivative of $ U $ in the interaction picture}\label{Ap:C}
In the interaction picture, $ U(t) = e^{iH_0t}e^{-iHt} $. Differentiate and use \eqref{Eq:4.1.1} and \eqref{Eq:4.1.6}
\begin{equation}\label{Eq:C.1}
\begin{split}
\dot{U}(t) &= ie^{iH_0t}H_0e^{-iHt} - ie^{iH_0t}He^{-iHt}\\
&= ie^{iH_0t}H_{\mathrm{int}}e^{-iHt}\\
&= ie^{iH_0t}H_{\mathrm{int}}e^{-iH_0t}e^{iH_0t}e^{-iHt}\\
&= -iH_{\mathrm{I}}(t)U(t).
\end{split}
\end{equation}

\section{Ehrenfest's theorem}\label{Ap:D}
\textbf{Theorem:} For a Hermitian operator $ A $
\begin{equation}\label{Eq:D.1}
\partial_0\langle A \rangle =\langle \partial_0 A \rangle - i \langle [ A, H]\rangle.
\end{equation}
\textit{Proof:} Differentiate $ \langle A \rangle $ using the product rule,
\begin{equation}\label{Eq:D.2}
\begin{split}
\partial_0 \langle f | A | f \rangle &= \langle f |\overleftarrow{\partial_0} A | f \rangle + \langle f | (\partial_0 A) | f \rangle + \langle f | A \partial_0 | f \rangle \\
&= i\langle f |HA | f \rangle + \langle f | (\partial_0 A) | f \rangle -i \langle f | A H | f \rangle\\
&=\langle \partial_0 A \rangle - i\langle [ A, H]\rangle,
\end{split}
\end{equation}
since $\partial_0 | f \rangle = - iH |f\rangle$ (the Schr\"{o}dinger equation), and since $ H $ is Hermitian.\\
\\
\textbf{Corollary:} For an observable quantity, $ A $, with no explicit time dependence, 
\begin{equation}\label{Eq:D.5}
\partial_0\langle A \rangle = - i \langle [ A, H]\rangle.
\end{equation}
\textbf{Theorem:} For the space indices, $a \in \lbrace 1, 2, 3\rbrace$, $\partial_a \langle A \rangle =  \langle \partial_a A \rangle  $.\\
\\
\textit{Proof:} Space translation is the same for an observable operator, $ A(x) $, and the corresponding classical observable, $ A_{\mathrm{c}}(x)= \langle A \rangle $. Hence, differentiating from first principles, 
\begin{equation}\label{Eq:D.6}
\begin{split}
\partial_a \langle A(x) \rangle &= \lim\limits_{dx^a\rightarrow 0}\frac{\langle A(x+dx)\rangle - \langle A(x)\rangle}{dx^a} \\
&= \lim\limits_{dx^a\rightarrow 0}\frac{\langle A(x+dx) -  A(x)\rangle}{dx^a} \\
&=\langle \partial_a A(x) \rangle.
\end{split}
\end{equation}

\section{Solution of the Dirac equation}\label{Ap:E}
The positive energy solutions to the Dirac equation are
\begin{equation}\label{Eq:E.1}
f_{\mu}(x) = \langle x,\mu|f \rangle = (\tfrac{1}{2\pi})^{3/2} \sum\limits_{r=1}^{2} \int d^3\boldsymbol{p}\, F(\boldsymbol{p},r)u_{\mu}(\boldsymbol{p},r) e^{-ix \cdot p},
\end{equation}
where $ p $ satisfies the mass shell condition and $ u $ is a Dirac spinor with the form of \eqref{Eq:3.3.7}\\
\\
\textit{Proof:} Observe that
\begin{equation}\label{Eq:E.2}
\boldsymbol{\sigma}\cdot\boldsymbol{p} = \begin{bmatrix} p^3 & p^1-ip^2\\ p^1+ip^2 & -p^3 \end{bmatrix}.
\end{equation}
So,
\begin{equation}\label{Eq:E.3}
\begin{split}
(\boldsymbol{\sigma}\cdot\boldsymbol{p})^2 &= 
\begin{bmatrix} p^3 & p^1-ip^2\\ p^1+ip^2 & -p^3 \end{bmatrix}
\begin{bmatrix} p^3 & p^1-ip^2\\ p^1+ip^2 & -p^3 \end{bmatrix}\\
&= \begin{bmatrix} (p^3)^2 + (p^1)^2+ (p^2)^2 & 0\\ 0 & (p^3)^2 + (p^1)^2+ (p^2)^2 \end{bmatrix}\\
&= ((p^0)^2 - m^2) 1_2.
\end{split}
\end{equation}
Hence,
\begin{equation} \label{Eq:E.4}
\begin{split}
p_a \gamma_{\mu\nu}^a u_{\nu}(\boldsymbol{p},r) 
&= \sqrt{\frac{p^0+m}{2p^0}}
\begin{bmatrix}
p^01_2 & - \boldsymbol{\sigma}\cdot\boldsymbol{p}\\
\boldsymbol{\sigma}\cdot\boldsymbol{p} & -p^01_2
\end{bmatrix}
\begin{bmatrix}
\zeta(r)\\
\frac{\boldsymbol{\sigma}\cdot\boldsymbol{p}}{p^0+m} \zeta(r)
\end{bmatrix} \\
&= \sqrt{\frac{p^0+m}{2p^0}} \begin{bmatrix}
( p^01_2 - \frac{ (\boldsymbol{\sigma}\cdot\boldsymbol{p})^2}{p^0+m})\zeta(r)\\
\boldsymbol{\sigma}\cdot\boldsymbol{p}\frac{p^0+m - p^0}{p^0+m} \zeta(r)
\end{bmatrix} \\
&= \sqrt{\frac{p^0+m}{2p^0}} \begin{bmatrix}
( p^01_2 - (p^0 - m) 1_2)\zeta(r)\\
\boldsymbol{\sigma}\cdot\boldsymbol{p}\frac{m }{p^0+m} \zeta(r)
\end{bmatrix}\\
&= m\sqrt{\frac{p^0+m}{2p^0}} \begin{bmatrix}
\zeta(r)\\
\frac{\boldsymbol{\sigma}\cdot\boldsymbol{p}}{p^0+m} \zeta(r)
\end{bmatrix}\\
&=mu_{\mu}(\boldsymbol{p},r).
\end{split}
\end{equation}
Then differentiation of \eqref{Eq:E.1} gives,
\begin{equation} \label{Eq:E.5}
\begin{split}
i\partial_a \gamma_{\mu\nu}^a f_{\nu}(x) 
&= (\tfrac{1}{2\pi})^{3/2}\sum\limits_{r=1}^{2} \int d^3\boldsymbol{p}\, p_a\gamma_{\mu\nu}^a F(\boldsymbol{p},r)u_{\nu}(\boldsymbol{p},r) e^{-ix \cdot p}\\
&= (\tfrac{1}{2\pi})^{3/2} \sum\limits_{r=1}^{2} \int d^3\boldsymbol{p}\, m F(\boldsymbol{p},r)u_{\mu}(\boldsymbol{p},r) e^{-ix \cdot p}\\
&=mf_{\mu}(x),
\end{split}
\end{equation}
as required. The analysis is similar for antiparticles.

\section{Normalisation of Dirac spinors}\label{Ap:F}
The Pauli spin matrices are Hermitian. So, using \eqref{Eq:E.3},
\begin{equation}\label{Eq:F.1}
\boldsymbol{\sigma}\cdot\boldsymbol{p}^{\dagger}\boldsymbol{\sigma}\cdot\boldsymbol{p} = (\boldsymbol{\sigma}\cdot\boldsymbol{p})^2 = ((p^0)^2 - m^2) 1_2.
\end{equation}
Then
\begin{equation} \label{Eq:F.2}
\begin{split}
u(\boldsymbol{p},r)^{\dagger} u(\boldsymbol{p},s)
&= \frac{p^0+m}{2p^0} \begin{bmatrix}
\zeta(r)^{\dagger} &
\zeta(r)^{\dagger}\frac{\boldsymbol{\sigma}\cdot\boldsymbol{p}}{p^0+m}^{\dagger}
\end{bmatrix}
\begin{bmatrix}
\zeta(r)\\
\frac{\boldsymbol{\sigma}\cdot\boldsymbol{p}}{p^0+m} \zeta(r)
\end{bmatrix}\\
&= \frac{p^0+m}{2p^0}(1+\frac{p^0-m}{p^0+m})\delta_{rs}\\
&=\delta_{rs}.
\end{split}
\end{equation}

\section{Integral form of the perturbation expansion}\label{Ap:G}

Integrate \eqref{Eq:C.1} directly,
\begin{equation} \label{Eq:G.1}
U(t)=1-i\int_{0}^{t}dt_1\,H_{\mathrm{I}}(t_1)U(t_1).
\end{equation}
Substituting $ U $ iteratively back into the integral gives the Dyson expansion,
\begin{equation} \label{Eq:G.2}
U(t)=1 + (-i)\int_{0}^{t}dt_1H_{\mathrm{I}}(t_1) +(-i)^2\int_{0}^{t}dt_1\int_{0}^{t_1}dt_2H_{\mathrm{I}}(t_1)H_{\mathrm{I}}(t_2) + \ldots.
\end{equation}
This can also be verified by differentiating. Each term is the derivative of the next multiplied by $ -iH(t) $. Substituting
\begin{equation} \label{Eq:G.3}
H_{\mathrm{I}}(x_i^0)=\chi^3 \sum\limits_{\mathrm{D}}I(x_i
) \approx \int d^3x_i\, I(x_i)
\end{equation}
gives
\begin{equation} \label{Eq:G.4}
U(t)\approx 1+ \sum\limits_{n\ge 1}(-i)^n \int 
d^4x_1  \int\limits_{x_2^0 < x_1^0}d^4x_2
\ldots \int\limits_{x_n^0 < x_{n-1}^0 }d^4x_n \,I(x_1) 
\ldots I(x_n).
\end{equation}
It can be seen that, provided that the integrals are defined,
\begin{multline*} 
\int d^4x_1 \int\limits_{x_2^0 < x_1^0}d^4x_2
\ldots \int\limits_{x_n^0 < x_{n-1}^0 }d^4 x_n\, I(x_1) 
\ldots I(x_n) \\
= \frac{1}{n!} \int d^4x_1 \int d^4x_2
\ldots \int d^4x_n \mathcal{T}\lbrace I(x_1) 
 \ldots I(x_n) \rbrace .
\end{multline*}
Hence, we can write the perturbation expansion
\begin{equation} \label{Eq:G.5}
U(t)\approx 1 + \sum\limits_{n\ge 1}\frac{(-i)^n}{n!} \int d^4x_1 
\ldots \int d^4x_n \mathcal{T}\lbrace I(x_1) 
 \ldots I(x_n) \rbrace .
\end{equation}

\section{Locality of Dirac field operators}\label{Ap:H}
\textbf{Theorem:} The equal time anticommutation relations for the Dirac field and Dirac adjoint and obey:
\begin{equation} \label{Eq:H.1}
\lbrace \psi_{\alpha}(x),\hat{\psi}_{\beta}(y) \rbrace_{x^0=y^0} = \gamma_{\alpha\beta}^0\delta(x-y).
\end{equation}
\textit{Proof:} Using $ (\boldsymbol{\sigma}\cdot\boldsymbol{p})^2 = ((p^0)^2 - m^2) 1_2 $ \eqref{Eq:E.3} and $ \sum_{r} \zeta(r)\zeta(r)^{\dagger} = 1_2$  (true in a particular basis, so true in any basis),
\begin{equation} \label{Eq:H.2}
\begin{split}
\sum\limits_{r} u(\boldsymbol{p},r) \hat{ u}(\boldsymbol{p},s)
&= \frac{p^0+m}{2p^0} \sum\limits_{r}
\begin{bmatrix}
\zeta(r)\\
\frac{\boldsymbol{\sigma}\cdot\boldsymbol{p}}{p^0+m} \zeta(r)
\end{bmatrix}
\begin{bmatrix}
\zeta(r)^{\dagger} &
-\zeta(r)^{\dagger}\frac{\boldsymbol{\sigma}\cdot\boldsymbol{p}}{p^0+m}^{\dagger}
\end{bmatrix}\\
&= \frac{1}{2p^0}\begin{bmatrix}
(p^0+m)1_2 & -\boldsymbol{\sigma}\cdot\boldsymbol{p}^{\dagger}\\
\boldsymbol{\sigma}\cdot\boldsymbol{p} & -\frac{(\boldsymbol{\sigma}\cdot\boldsymbol{p})^2}{p^0+m}1_2
\end{bmatrix}\\
&= \frac{1}{2p^0}\begin{bmatrix}
(p^0+m)1_2 & -\boldsymbol{\sigma}\cdot\boldsymbol{p}\\
\boldsymbol{\sigma}\cdot\boldsymbol{p} & -(p^0-m)1_2
\end{bmatrix}\\
&=\frac{1}{2p^0}(p \cdot \gamma + m).
\end{split}
\end{equation}
Similarly,
\begin{equation} \label{Eq:H.3}
\sum\limits_{r} v(\boldsymbol{p},r) \hat{ v}(\boldsymbol{p},s) =\frac{1}{2p^0}(p \cdot \gamma - m).
\end{equation}
We have 
\begin{equation} \label{Eq:H.4}
\lbrace \psi_{\alpha}(x),\hat{\psi}_{\beta}(y) \rbrace = \langle \underline{x},\alpha | \hat{\underline{y}}, \beta \rangle + \langle \hat{\bar{y}},\beta | \bar{x}, \alpha \rangle ^{\mathsf{T}},
\end{equation}
where $ ^{\mathsf{T}} $ denotes that $ \alpha $ and $ \beta $ are transposed. Using the resolution of unity and the solution of the Dirac equation,
\begin{equation} \label{Eq:H.5}
\begin{split}
\langle \underline{x}, \alpha | \hat{\underline{y}}, \beta \rangle 
&= \frac{1}{8\pi^3}\sum_r \int d^3\boldsymbol{p}\, u_{\alpha}(\boldsymbol{p},r) \hat{ u}_{\beta}(\boldsymbol{p},r) e^{-ip \cdot (x-y)} \\
&= \frac{1}{8\pi^3} \int \frac{d^3\boldsymbol{p}}{2p^0} \,(p\cdot\gamma +m)_{\alpha\beta} e^{-ip \cdot (x-y)}.
\end{split}
\end{equation}
Likewise for the antiparticle,
\begin{equation} \label{Eq:H.6}
\langle \hat{\bar{y}},\beta | \bar{x}, \alpha \rangle ^{\mathsf{T}}
= \frac{1}{8\pi^3} \int \frac{d^3\boldsymbol{p}}{2p^0} \,(p\cdot\gamma -m)_{\alpha\beta} e^{ip \cdot (x-y)}.
\end{equation}
Substituting $ \boldsymbol{p}\rightarrow - \boldsymbol{p} $ at  $ x^{0} = y^{0}$,
\begin{equation} \label{Eq:H.7}
\langle \hat{\bar{y}},\beta | \bar{x}, \alpha \rangle ^{\mathsf{T}}
= \frac{1}{8\pi^3} \int \frac{d^3\boldsymbol{p}}{2p^0}\, (2p_0\gamma^0 - p\cdot\gamma -m)_{\alpha\beta} e^{-ip \cdot (x-y)}.
\end{equation}
Adding \eqref{Eq:H.5} and \eqref{Eq:H.7} at $ x^{0} = y^{0} $ gives the equal time anticommutator, \eqref{Eq:H.4},
\begin{equation} \label{Eq:H.8}\lbrace \psi_{\alpha}(x),\hat{\psi}_{\beta}(y) \rbrace_{x^{0} = y^{0}} =  \frac{\gamma_{\alpha\beta}^0}{8\pi^3} \int d^3\boldsymbol{p}\, e^{-ip \cdot (x-y)} = \gamma_{\alpha\beta}^0 \delta(\boldsymbol{x}-\boldsymbol{y}).
\end{equation}
\textbf{Theorem:} The anticommutation relation for the Dirac field and the Dirac adjoint is zero outside the light cone.\\
\\
\textit{Proof:} From \eqref{Eq:H.5},
\begin{equation} \label{Eq:H.9}
\langle \underline{x}, | \hat{\underline{y}}, \rangle = \frac{1}{8\pi^3}(i\partial\cdot\gamma +m) \int \frac{d^3\boldsymbol{p}}{2p^0}\, e^{-ip \cdot (x-y)},
\end{equation}
and from \eqref{Eq:H.6}
\begin{equation} \label{Eq:H.10}
\langle \hat{\bar{y}} | \bar{x} \rangle ^{\mathsf{T}}
= -\frac{1}{8\pi^3}(i\partial\cdot\gamma +m) \int \frac{d^3\boldsymbol{p}}{2p^0}\,  e^{ip \cdot (x-y)}.
\end{equation}
The anticommutator is found by adding:
\begin{equation} \label{Eq:H.11}
\begin{split}
\lbrace \psi_{\alpha}(x),\hat{\psi}_{\beta}(y) \rbrace 
&= \frac{1}{8\pi^3}(i\partial\cdot\gamma +m) \int \frac{d^3\boldsymbol{p}}{2p^0}\, (e^{-ip \cdot (x-y)}-e^{ip \cdot (x-y)})\\
&= \frac{1}{8\pi^3}(i\partial\cdot\gamma +m) \int d^4p \, (e^{-ip \cdot (x-y)}-e^{ip \cdot (x-y)})\delta(p^2-m^2)\\
\end{split}
\end{equation}
using the generalised scaling property of the delta function applied to the mass shell condition. The integral is Lorentz invariant and is zero when $ x^{0} - y^{0} = 0 $. We conclude that it is zero whenever $ x - y $ is spacelike.

\section{Gauge invariance}\label{Ap:I}
The local phase transformation, $ \psi(x) \rightarrow e^{i\alpha(x)}\psi(x), $ applied to the field operators, makes no difference to the current and so leaves the predictions of the theory unchanged (equivalently the transformation may be applied to the creation operators, remembering the sign change for antiparticles). The interacting Dirac equation, \eqref{Eq:6.2.7}, 
\begin{equation}\label{Eq:I.1}
(\gamma^a(i\partial_a - e \langle A_a(x) \rangle) -m)f(x)= 0,
\end{equation}
can be written, in terms of creation operators acting on any ket $ |f\rangle $,
\begin{equation}\label{Eq:I.2}
\int d^3x\, |\underline{x}\rangle(\gamma^a(i\partial_a - e \langle A_a(x) \rangle) -m)\langle \underline{x}| |f\rangle = 0.
\end{equation}
A local gauge transformation applied to the creation operators, $|\underline{x}\rangle \rightarrow e^{-i\alpha(x)}|\underline{x}\rangle $, gives
\begin{equation}\label{Eq:I.3}
\int d^3x\, |\underline{x}\rangle e^{-i\alpha(x)}(\gamma^a(i\partial_a - e \langle A_a(x) \rangle) -m)e^{i\alpha(x)}\langle \underline{x}| |f\rangle = 0.
\end{equation}
So,
\begin{equation*}
\int d^3x\, |\underline{x}\rangle (\gamma^a(i\partial_a -\partial_a\alpha(x) - e \langle A_a(x) \rangle) -m)\langle \underline{x}| |f\rangle = 0.
\end{equation*}
\begin{equation}\label{Eq:I.4}
\int d^3x\, |\underline{x}\rangle (\gamma^a(i\partial_a  - e \langle A_a(x) \rangle') -m)e^{i\alpha(x)}\langle \underline{x}| |f\rangle = 0.
\end{equation}
which is identical to \eqref{Eq:I.1} apart from the replacement 
\begin{equation}\label{Eq:I.5}
\langle A_a(x) \rangle' = \langle A_a(x) \rangle + e^{-1}\partial_a\alpha(x).
\end{equation}
But the Faraday tensor \eqref{Eq:6.3.8} 
is also unchanged by this replacement;
\begin{equation}\label{Eq:I.6}
F^{ab} = \partial^a \langle A^b(x) \rangle - \partial^b \langle A^a(x) \rangle = \partial^a \langle A^b(x) \rangle ' - \partial^b \langle A^a(x) \rangle '.
\end{equation}
So the local phase symmetry of the field operators is precisely equivalent to the well known gauge symmetry of the classical electromagnetic field.

\section{Feynman diagrams}\label{Ap:J}

\subsection{The time-ordered vertex for qed}\label{Ap:J.1}
The interaction density for qed is given by \eqref{Eq:6.1.1},
\begin{equation}\label{Eq:J.1}
I(x)=ej^a(x)A_a(x) = e(|\hat{\underline{x}}\rangle + \langle\hat{\bar{x}}|) \gamma^a (|\bar{x}\rangle + \langle \underline{x}|) (|\underline{x},a\rangle + \langle \underline{x},a| ),
\end{equation}
where photon creation and annihilation operators are distinguished by the vector index, $ a $. $ I(x) $ is the sum of eight terms, each of which can be represented can be represented diagrammatically as a time-ordered vertex or node (figure 3). Lines above the node correspond to creation operators, and those below the node correspond to annihilation operators. The photon is represented by a wavy line, electrons by a upward arrow and positrons by a downward arrow. 
\begin{figure}[ht]
	\centering
		\includegraphics[width=0.5\textwidth]{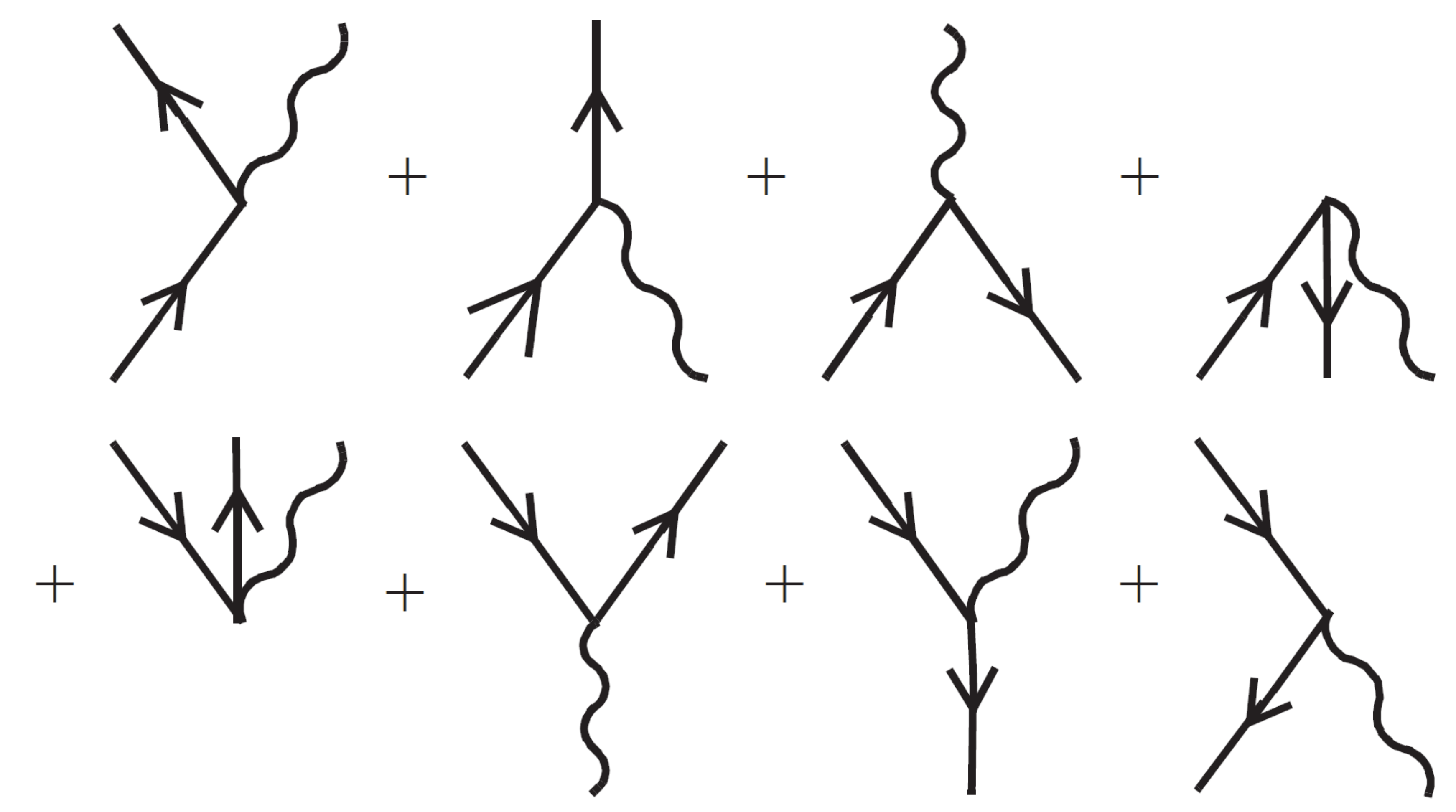}
	\caption{Time-ordered diagrams for qed.}
	\label{Fig:3}
\end{figure}
\subsection{Wick's theorem}\label{Ap:J.2}
Wick's theorem can be used to replace the time-ordered product with a normal ordered product by (anti)commuting annihilation operators to the right and creation operators to the left. Let $ \phi=|\bar{x}\rangle + \langle\underline{x}| $ be a field operator. If $ x^{0} < y^{0} $, the Feynman propagator, $ D(x - y) $, \eqref{Eq:7.1.1}, gives the amplitude for the creation of an antiparticle at $ x $ and its annihilation at $ y $. If $ x^{0} > y^{0} $ it gives the amplitude for creation of a particle at $ y $ and its annihilation at $ x $.\\
\\
\textbf{Theorem:} (Wick's Theorem) For two field operators,
\begin{equation}\label{Eq:J.2.1}
\mathcal{T}\lbrace \phi^{\dagger}(x) \phi(y) \rbrace = :\phi^{\dagger}(x) \phi(y) : + D(x-y).
\end{equation}
For $ n $ field operators:
\begin{multline}\label{Eq:J.2.2}\mathcal{T}\lbrace \phi^{\dagger}(x_1) \ldots \phi^{\dagger}(x_i) \phi(x_{i+1}) \ldots \phi(x_{n}) \rbrace\\ = :\phi^{\dagger}(x_1) \ldots \phi^{\dagger}(x_i) \phi(x_{i+1}) \ldots \phi(x_{n}) : \\ 
+ \sum_{\substack{\text{all pairs of} \\ 
\text{contractions}}}
  :\phi^{\dagger}(x_1) \ldots \phi(x_{n}) : \prod\limits_{\mathrm{pairs}\, j,k}D(x_j-x_k),
\end{multline}
where contracted pairs are omitted in the normal ordered product under the sum.\\
\\
A detailed proof by induction can be carried out, but the proof is no more evident than the theorem itself, which just means that we do the normal ordering by carrying out the contractions.

\subsection{The S-matrix}\label{Ap:J.3}

Initial and final kets can be expressed as sums of plane wave kets by using the resolution of unity in momentum space. The time evolution between $ t_{0} $ and $ t_{1} $ is given by a matrix in momentum space $ \langle p_1;\ldots;p_j|U(t_1,t_0)|p_{j+1};\ldots;p_n\rangle $. In scattering experiments, the initial ket (generated by a particle accelerator), and the final ket (typically measured by bubble chamber, wire chamber or silicon detector) are well represented as pure momentum kets. In this case the interesting interaction takes place at the scattering event, and $ t_{0} $ and $ t_{1} $ are not important.\\
\\
\textbf{Postulate:} The \textbf{S-matrix} (or \textbf{scattering matrix}) is 
\begin{equation}\label{Eq:J3.1}
\langle p_1;\ldots;p_j|S|p_{j+1};\ldots;p_n\rangle = \lim_{\substack{t_0\rightarrow -\infty \\ t_1\rightarrow \infty}} \langle p_1;\ldots;p_j|U(t_1,t_0)|p_{j+1};\ldots;p_n\rangle.
\end{equation}
?
The $ S $-matrix is found from the perturbation expansion by first normal ordering the terms using Wick's theorem. Then, for the interaction density at $ x $, the creation operator acting on the initial ket $ |\boldsymbol{p},r\rangle $ gives, for a photon,
\[ \langle \underline{x}|\boldsymbol{p},r \rangle = (\tfrac{1}{2\pi})^{3/2} \frac{w(\boldsymbol{p},r)}{\sqrt{2p^0}}e^{-ip \cdot x}, \]
for a Dirac particle,
\[ \langle \underline{x}|\boldsymbol{p},r \rangle = (\tfrac{1}{2\pi})^{3/2} u(\boldsymbol{p},r)e^{-ip \cdot x}, \]
and for an antiparticle,
\[ \langle \hat{\bar{x}}|\boldsymbol{p},r \rangle = (\tfrac{1}{2\pi})^{3/2} \hat{v}(\boldsymbol{p},r)e^{-ip \cdot x}. \]

Similarly, the annihilation operators in the interaction density acting on the final ket  gives, for a photon, 
\[ \langle \boldsymbol{p},r |\underline{x} \rangle = (\tfrac{1}{2\pi})^{3/2} \frac{w(\boldsymbol{p},r)}{\sqrt{2p^0}}e^{ip \cdot x}, \]
for a Dirac particle,
\[ \langle \boldsymbol{p},r | \hat{\underline{x}} \rangle = (\tfrac{1}{2\pi})^{3/2} \hat{u}(\boldsymbol{p},r)e^{ip \cdot x}, \]
and for an antiparticle,
\[ \langle \boldsymbol{p},r | \bar{x} \rangle = (\tfrac{1}{2\pi})^{3/2} v(\boldsymbol{p},r)e^{ip \cdot x}. \]
\begin{figure}[ht]
	\centering
		\includegraphics[width=0.5\textwidth]{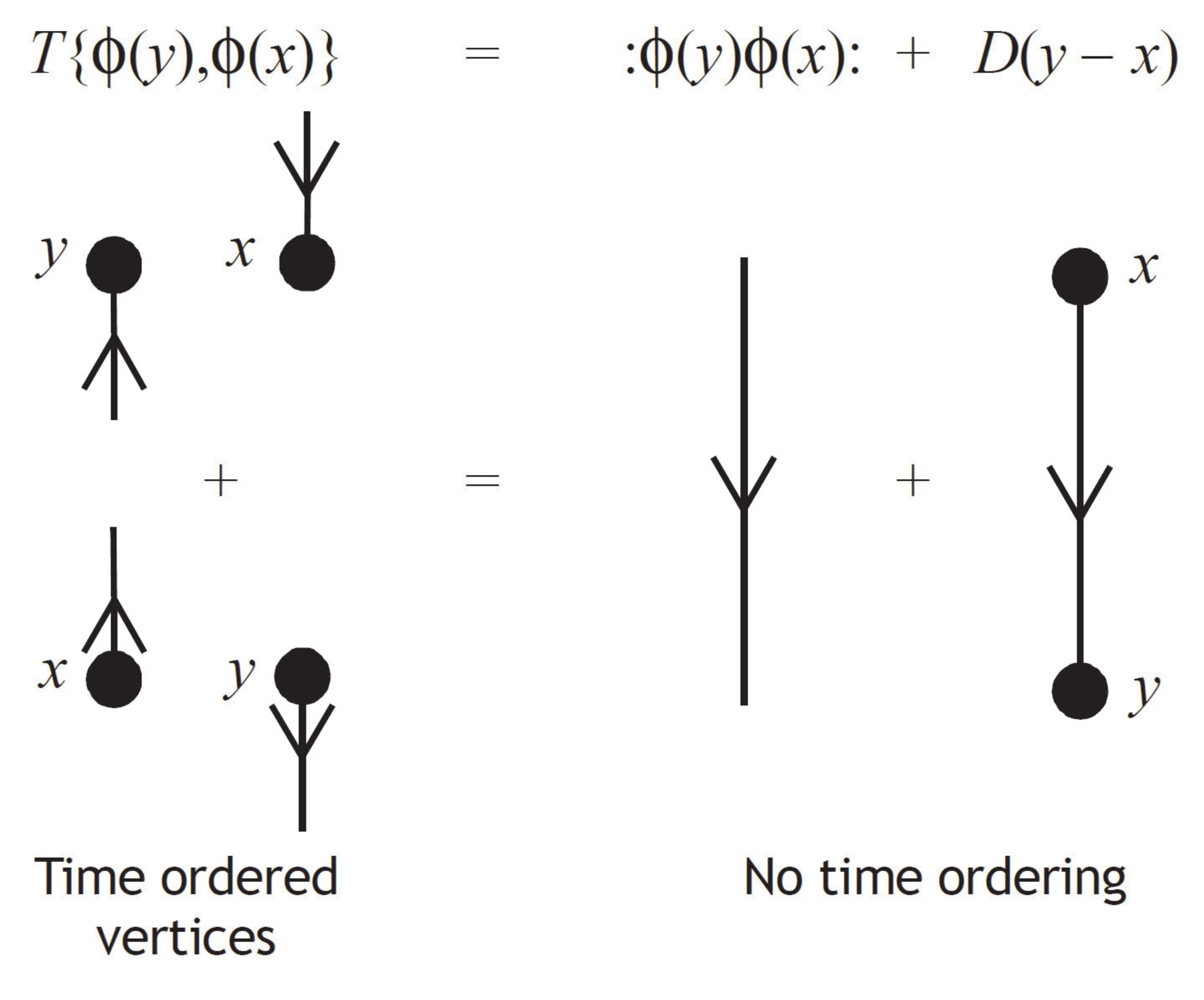}
	\caption{Contraction represented by connecting vertices and removing time-ordering.}
	\label{Fig:4}
\end{figure}

To keep track of the contractions in normal ordering the perturbation expansion, the terms are represented by graphs. A particle created at $ x^{0} $ may be annihilated at a later time $ y^{0} $. An antiparticle created at $ x^{0} $ may be annihilated at an earlier time $ y^{0} $. Each contraction is represented by connecting the corresponding lines between vertices, and, at the same time, removing time-ordering (figure 4). After carrying out the contractions, all topologically equivalent time-ordered diagrams are combined into a single diagram with no time-ordering between the nodes (figure 5). There are $ k! $ diagrams with $ k $ nodes. So, removing the ordering of nodes generates a factor $ k! $ and cancels the factor $ 1 / k! $ in the perturbation expansion \eqref{Eq:4.5.2}, leaving a sum for a diagram with $k $ vertices,
\[ U(t_n) = (-ie)^k \gamma^{a_1}\gamma^{a_2} \ldots \gamma^{a_n} (-i\chi^4)^k \sum\limits_{i_k \ne i_{k-1},\ldots ,i_1} \ldots \sum\limits_{i_2\ne i_1} \sum\limits_{i_1=1}^{n}.
\]
\begin{figure}[ht]
	\centering
		\includegraphics[width=0.5\textwidth]{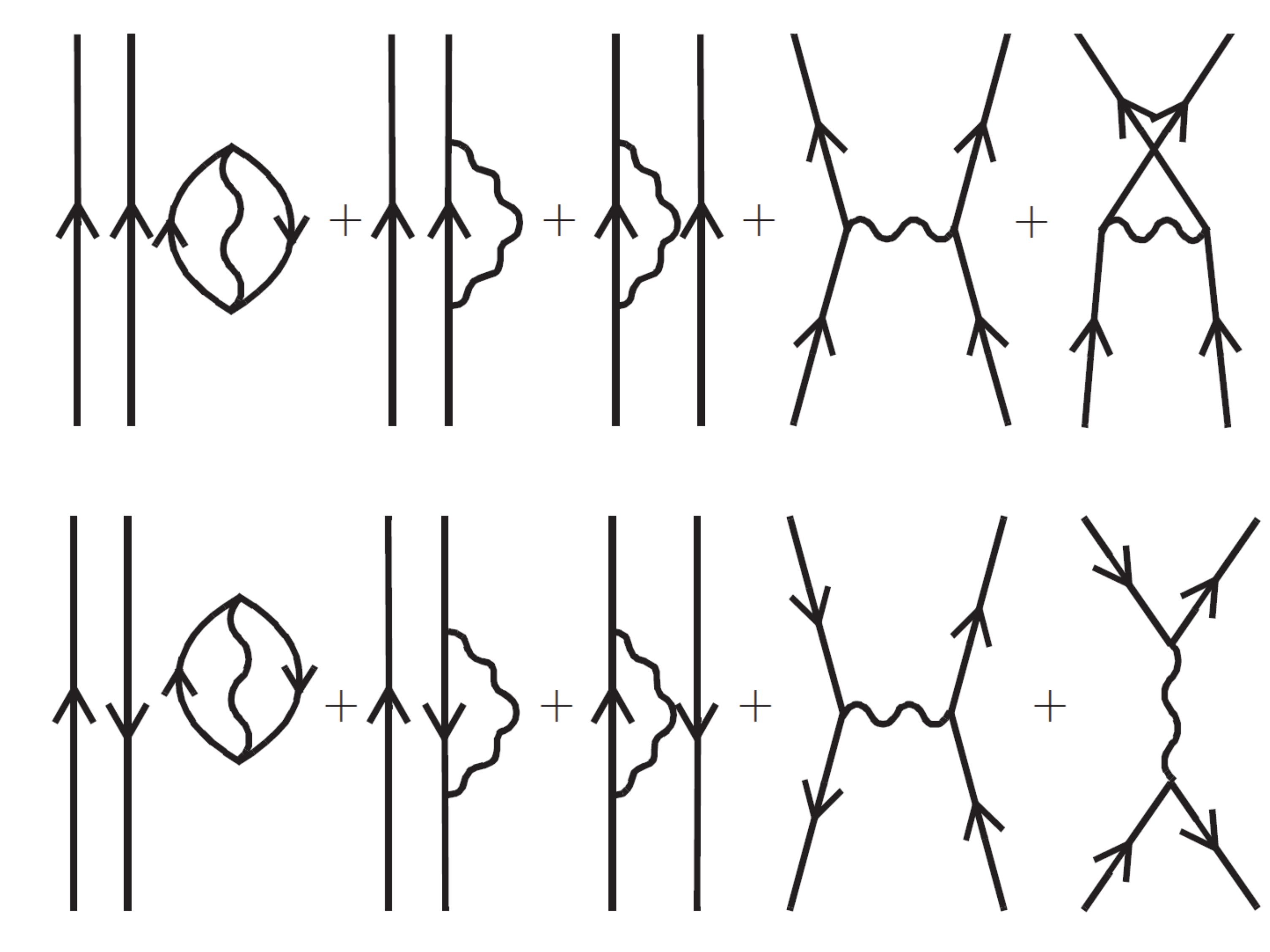}
	\caption{Second order diagrams for initial and final states with two particles (top), and with a particle and antiparticle (bottom). The final term of the top diagram is zero if the particles can be distinguished (e.g. if one is bound to an atom).}
	\label{Fig:5}
\end{figure}

\subsection{Conservation of energy and momentum}\label{Ap:J.4}

Gather all the exponential terms from internal and external lines with $ x_{i} $ in the exponent. Provided the time from $t_{0}$ to $ t_{n} $ is large, the result is a delta function,
\begin{equation}\label{Eq:J4.1}
(\frac{\chi}{2\pi})^4 \sum_{\mathrm{S}} e^{-i\tilde{p}\cdot x_i + i\tilde{q}\cdot x_i -i\tilde{k}\cdot x_i  } = \delta^{(4)}(\tilde{p}-\tilde{q}+\tilde{k})
\end{equation}
where $ \tilde{p}, \tilde{q}, \tilde{k} $ refer to the arrowed line coming from the vertex, the arrowed line going into the vertex, and the photon line, respectively. The delta function shows that the tilda'd quantities are conserved. For internal lines, $ \tilde{p}^0, \tilde{q}^0, \tilde{k}^0 $, are the dummy variables introduced in the contour integration. For external lines $(\tilde{p}^0, \tilde{q}^0, \tilde{k}^0) = ( p^0, q^0, k^0 )$. Energy, $ p^{0} $, was originally defined to be the zero component of a vector. This is not a conserved quantity. Vectors are products of measurement, and only have real meaning in measurement. By definition, internal lines do not correspond to measured states. So, $ p^{0} $ has no meaning on internal lines in a Feynman diagram. The conserved tilda'd quantities are of more interest than vector quantities and it is usual to \textit{redefine} energy.\\
\\
\textbf{Redefinition}: \textbf{Energy} is the conserved quantity, $ \tilde{p}^0 $, which appears on the lines of a Feynman diagram.\\
\\
With this definition, energy-momentum, $ \tilde{p} $, is conserved, but is not a vector, and does not obey the mass shell condition on internal lines in Feynman diagrams. Particles are said to be off shell on internal lines. On external lines, representing measured states, this definition of energy coincides with the original definition for measured states, as the time component of a vector. Particles are said to be on shell on external lines, meaning that the mass shell condition is obeyed in measurement.

\subsection{Feynman rules}\label{Ap:J.5}

After using the delta functions to carry out the integrals over tilda'd quantities, and imposing the rule that energy-momentum is conserved at each vertex, there remains an integral for each independent internal loop,
\[ \tfrac{1}{16\pi^4} \int d^4\tilde{p}\,.   \]
Each vertex contributes a factor
\[-ie\gamma^a. \]
For external lines in the initial state we have, for a photon,
\[ \sqrt{\tfrac{1}{4\pi p^0} } w(\boldsymbol{p},r), \]
for a Dirac particle,
\[ \sqrt{\tfrac{1}{2\pi} } u(\boldsymbol{p},r), \]
and for an antiparticle,
\[ \sqrt{\tfrac{1}{2\pi} } \hat{v}(\boldsymbol{p},r). \]
For external lines in the final state we have, for a photon,
\[ \sqrt{\tfrac{1}{4\pi p^0} } w(\boldsymbol{p},r), \]
for a Dirac particle,
\[ \sqrt{\tfrac{1}{2\pi} } \hat{u}(\boldsymbol{p},r), \]
and for an antiparticle,
\[ \sqrt{\tfrac{1}{2\pi} }v(\boldsymbol{p},r). \]
For internal arrowed lines we have
\[ \frac{ip \cdot \gamma +m }{\tilde{p}^2 - m^2 + i\epsilon}, \]
and for internal photon lines we have
\[\frac{-ig_{ab}}{\tilde{p}^2 + 2i|\boldsymbol{p}| \epsilon +\epsilon^2} .\]
In addition there is a minus sign if an odd number of commutations of Fermion creation and annihilation operators is required to put the diagram into normal order. The limit $ \epsilon \rightarrow 0 $ should be taken after evaluation of integrals for loops and for the initial and final states. If $ |\boldsymbol{p}| > 0 $ then the photon propagator can be replaced with 
\[\frac{-ig_{ab}}{\tilde{p}^2  +\epsilon} .\]
Certain diagrams contain a divergence when photon energy goes to zero. In this case $ \epsilon^2 $ should be retained until after evaluation of the integral to control the infrared divergence ($ \epsilon^2 $ plays the role of the small photon mass commonly used for this purpose).

\section{Derivation of propagators}\label{Ap:K}
\subsection{Lemma}\label{Ap:K.1}
\begin{multline}\label{Eq:K.1.1}
\lim\limits_{\epsilon\rightarrow 0^+}\int_{-\infty}^{\infty}d\tilde{p}^0\,\frac{e^{-i\tilde{p}^0 x^0}}{\tilde{p}^2 - m^2 + 2ip^0 \epsilon + \epsilon^2} \\ =  -2\pi i \left[ \frac{e^{-ip^0x^0}}{2p^0} \Theta(x^0) +
\frac{e^{ip^0x^0}}{2p^0} \Theta(-x^0)
\right].
\end{multline}
where $ \Theta(t) = 0 $ if $ t\le 0 $, $ \Theta(t) = 1 $ if $t > 0$.\\
\\
\textit{Proof:} Since $ \tilde{p}= (\tilde{p}^0, \pm \boldsymbol{p}) $,
\[\tilde{p}^2 -m^2 = \tilde{p}^2 -p^2 = (\tilde{p}^0)^2 - (p^0)^2. \]
So,
\begin{equation}\label{Eq:K.1.2}
\begin{split}
\int_{-\infty}^{\infty}d\tilde{p}^0\,\frac{e^-i\tilde{p}^0 x^0}{\tilde{p}^2 - m^2 + 2ip^0 \epsilon + \epsilon^2} 
&= \int_{-\infty}^{\infty}d\tilde{p}^0\,\frac{e^-i\tilde{p}^0 x^0}{(\tilde{p}^0)^2 - (p^0 - i\epsilon)^2}\\
&= \int_{-\infty}^{\infty}d\tilde{p}^0\,\frac{e^-i\tilde{p}^0 x^0}{(\tilde{p}^0-p^0 + i\epsilon)(\tilde{p}^0+p^0 - i\epsilon)}.
\end{split}
\end{equation}
This is evaluated as a contour integral, noting that the integral on $ C_{2} $ vanishes in the lower half plane if $ x^0>0 $, and in the upper half plane if $ x^0<0 $ (figure 6). The integral does not exist if $x^0=0$.
\begin{figure}[ht]
	\centering
		\includegraphics[width=1\textwidth]{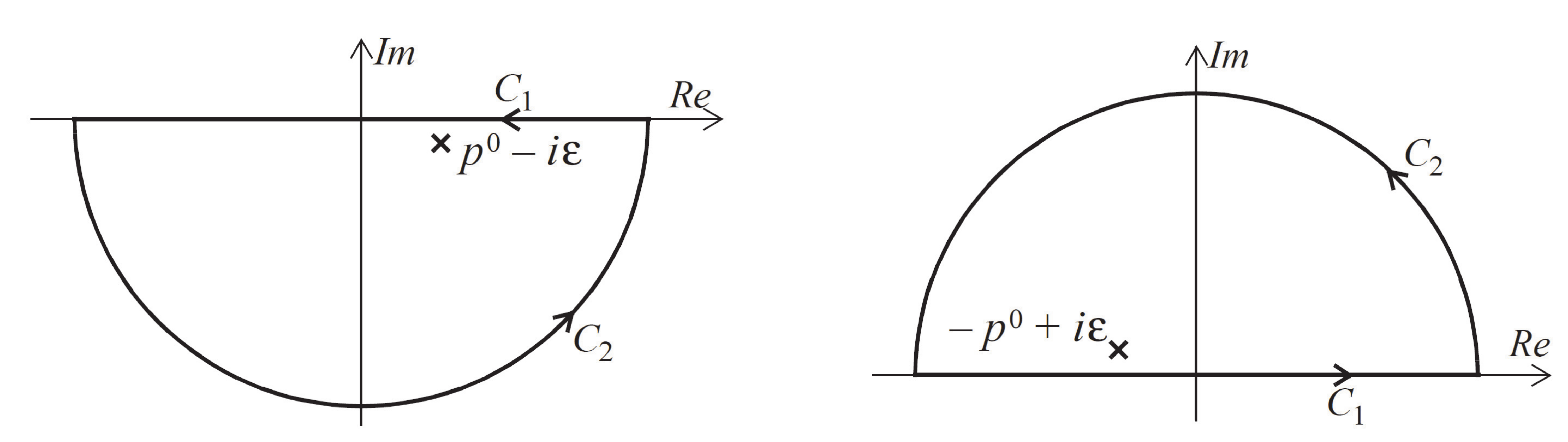}
	\caption{Contours for the integrations in the complex $ \tilde{p}^0 $ plane.}
	\label{Fig:6}
\end{figure}
\subsection{The photon propagator}\label{Ap:K.2}
Using the lemma \eqref{Eq:K.1.1},
\begin{equation*}
\begin{split}
D_F(x-y)&=\Theta(x^0-y^0)\langle \underline{x}|\underline{y} \rangle + \Theta(y^0-x^0)\langle \underline{y}|\underline{x} \rangle^{\mathsf{T}} \\
&=\frac{-g}{8\pi^3}\int d^3\boldsymbol{p}\, \left[  \Theta(x^0 - y^0)\frac{e^{-ip \cdot (y-x)}}{2p^0} + \Theta(y^0 - x^0)\frac{e^{ip \cdot (y-x)}}{2p^0} \right]  \\
&=\frac{-ig}{16\pi^4} \lim_{\epsilon \rightarrow 0^+}\int d^3\tilde{\boldsymbol{p}} \int_{-\infty}^{\infty} d\tilde{p}^0 \, \lbrack \Theta(x^0 - y^0) + \Theta(y^0 - x^0) \rbrack \frac{e^{-i\tilde{p} \cdot (y-x)}}{\tilde{p}^2 + 2ip^0 \epsilon + \epsilon^2},
\end{split}
\end{equation*}
where the energy-momentum from $ y $ to $ x $ is $ \tilde{p}=(\tilde{p}^0,\boldsymbol{p}) $ if $ x^0>y^0 $ or $ \tilde{p}=(\tilde{p}^0,-\boldsymbol{p}) $ if  $ y^0>x^0 $. If $ y^0\ne x^0 $, the step functions can be summed to unity, 
\begin{equation}\label{Eq:K.2.1}
D_F(x-y)=\frac{-ig}{16\pi^4} \lim_{\epsilon \rightarrow 0^+}\int d^4p \, \frac{e^{-i\tilde{p} \cdot (y-x)}}{\tilde{p}^2 + 2ip^0 \epsilon + \epsilon^2}.
\end{equation}
\subsection{The Dirac propagator}\label{Ap:K.3}
Using lemma \eqref{Eq:K.1.1} with \eqref{Eq:H.9} and \eqref{Eq:H.10},
\begin{equation*}
\begin{split}
S_F(x-y)&=\Theta(x^0-y^0)\langle \underline{x}|\hat{\underline{y}} \rangle - \Theta(y^0-x^0)\langle \hat{\bar{y}}|\bar{y} \rangle^{\mathsf{T}} \\
&=\frac{i\partial \cdot \gamma + m}{8\pi^3}\int \frac{d^3\boldsymbol{p}}{2p^0}\, \lbrack \Theta(x^0 - y^0)e^{-ip \cdot (y-x)} + \Theta(y^0 - x^0)e^{ip \cdot (y-x)} \rbrack \\
&=i \lim_{\epsilon \rightarrow 0^+} \frac{i\partial \cdot \gamma + m}{16\pi^4} \int d^3\tilde{\boldsymbol{p}} \int_{-\infty}^{\infty} d\tilde{p}^0 \,\lbrack \Theta(x^0 - y^0) + \Theta(y^0 - x^0) \rbrack \frac{e^{-i\tilde{p} \cdot (y-x)}}{\tilde{p}^2 -m^2 + 2ip^0 \epsilon + \epsilon^2},
\end{split}
\end{equation*}
where the energy-momentum in the direction of the arrow from $ y $ to $ x $ is $ \tilde{p}=(\tilde{p}^0,\boldsymbol{p}) $ if $ x^0>y^0 $ or $ \tilde{p}=(\tilde{p}^0,-\boldsymbol{p}) $ if  $ y^0>x^0 $ . If $ y^0\ne x^0 $, the step functions can be summed to unity, 
\begin{equation}\label{Eq:K.3.1}
S_F(x-y)=i \lim_{\epsilon \rightarrow 0^+} \frac{i\partial \cdot \gamma + m}{16\pi^4} \int d^4\tilde{p}\, \frac{e^{-i\tilde{p} \cdot (y-x)}}{\tilde{p}^2 -m^2 + 2ip^0 \epsilon + \epsilon^2}. 
\end{equation}
For a Dirac particle, $ p^0>0 $, and we can simplify the denominator by shifting the pole under the limit, replacing $ 2ip^0\epsilon +\epsilon^2 $ with $ i\epsilon $. Thus the Dirac propagator arrowed from $ y $ to $ x $ is
\begin{equation}\label{Eq:K.3.2}
S_F(x-y)= \frac{i}{16\pi^4} \lim_{\epsilon \rightarrow 0^+} \int d^4\tilde{p}\, \frac{(\tilde{p} \cdot \gamma + m)e^{-i\tilde{p} \cdot (y-x)}}{\tilde{p}^2 -m^2 + i \epsilon }. 
\end{equation}
\section{The origin of the ultraviolet divergence}\label{Ap:L}
It is well known that an integral which contains a squared delta-function in the integrand,
\begin{equation}\label{Eq:L.1}
S = \int_a^b \delta^2(x-h) dx,
\end{equation}
cannot be defined for $ a<h<b $. The origin of the ultraviolet divergence has been identified as the inclusion of such terms in Feynman diagrams containing loop integrals. Now consider the improper integral with $ \epsilon > 0 $,
\begin{equation}\label{Eq:L.2}
S' = \lim_{\epsilon \rightarrow 0} \int_a^{h-\epsilon} \delta^2(x-h) dx + \lim_{\epsilon \rightarrow 0} \int_{h+\epsilon}^b \delta^2(x-h) dx. 
\end{equation}
$ S' $ is (or at least, can be) well defined and trivially evaluates to zero. When the order of taking limits is properly tracked, the origin of the ultraviolet divergence is seen in the replacement of well defined integrals containing terms of the form $ S' $ with undefined integrals containing terms of the form $ S $.

The usual method of subtracting divergent quantities from loop diagrams is then seen as equivalent to subtracting a term,
\begin{equation}\label{Eq:L.3}
S'' = S - S'. 
\end{equation}
This restores the correct answer, but it means working with undefined quantities and the usual rationale is incorrect. No renormalisation is involved, and nor does the subtraction require adding counter terms to the Hamiltonian, because when the order of taking limits is tracked the divergence is not present in the original form of the perturbation expansion.

The appearance of squared delta functions in the integrand can be traced to the equal point multiplication between fields, which cannot be defined when fields are operator valued distributions. However, there is no equal point multiplication in \eqref{Eq:G.4}, because all inequalities in the bounds of integration are strict. The equal point multiplication appears in \eqref{Eq:G.5} as a consequence of incorrectly changing the order of taking limits. The exclusion of the equal point multiplication can also be seen as a physical constraint, that an electron cannot interact more than once in any instant. This statement is given a clear physical meaning through the introduction of a minimum discrete unit of time, $ \chi $.

\end{document}